\documentclass[iop]{emulateapj}
\usepackage{longtable}
\usepackage{tablefootnote}
\usepackage{color}
\usepackage{subfigure}
\usepackage{graphicx}

\shorttitle{The APOGEE SDSS-III Radial Velocity Survey of M dwarfs I}
\shortauthors{Deshpande et al.}

\begin{document}


\title{The SDSS-III APOGEE Radial Velocity Survey of M dwarfs I: Description of Survey and Science Goals}

\author{R.~Deshpande\altaffilmark{1,2}, C.~H.~Blake\altaffilmark{3,4}, C.~F.~Bender\altaffilmark{1,2}, S.~Mahadevan\altaffilmark{1,2}, R.C.~Terrien\altaffilmark{1,2},  J.~Carlberg\altaffilmark{5}, G.~Zasowski\altaffilmark{6}, J.~Crepp\altaffilmark{7}, A.~S.~Rajpurohit\altaffilmark{8}, C.~Reyl\'e\altaffilmark{8,9}, D.~L.~Nidever\altaffilmark{9}, D.~P.~Schneider\altaffilmark{1,2}, C.~Allende~Prieto\altaffilmark{11,12}, D.~Bizyaev\altaffilmark{13}, G.~Ebelke\altaffilmark{13}, S.~W.~Fleming\altaffilmark{1,2}, P.~M.~Frinchaboy\altaffilmark{14}, J.~Ge\altaffilmark{15},  F.~Hearty\altaffilmark{6},  J.~I.~Gonz\'alez~Hern\'andez\altaffilmark{11,12},E.~Malanushenko\altaffilmark{13}, V.~Malanushenko\altaffilmark{13}, S.R.~Majewski\altaffilmark{6}, D.~Oravetz\altaffilmark{13}, K.~Pan\altaffilmark{13}, R.~P.~Schiavon\altaffilmark{16}, M.~Shetrone\altaffilmark{17}, A.~Simmons\altaffilmark{13}, K.~G.~Stassun\altaffilmark{18}, J.~C.,~Wilson\altaffilmark{6}, Wisniewski, J.~P.\altaffilmark{19}}


\altaffiltext{1}{Center for Exoplanets and Habitable Worlds, The Pennsylvania State University, University Park, PA 16802}
\altaffiltext{2}{Department of Astronomy and Astrophysics, The Pennsylvania State University, University Park, PA 16802}
\altaffiltext{3}{Department of Physics \& Astronomy, University of Pennsylvania, Philadelphia, PA 19104, USA}
\altaffiltext{4}{Department of Astrophysical Sciences, Princeton University, Peyton Hall, Ivy Lane, Princeton, NJ 08544, USA}
\altaffiltext{5}{Department of Terrestrial Magnetism, Carnegie Institution of Washington, 5241 Broad Branch Road NW, Washington, DC 20015, USA}
\altaffiltext{6}{University of Virginia, 530 McCormick Road, Charlottesville, VA 22904}
\altaffiltext{7}{University of Notre Dame}
\altaffiltext{8}{Institut UTINAM, CNRS UMR 6213, Observatoire des Sciences de l'Univers THETA Franche-Comt\'{e}-Bourgogne, Universit\'e de Franche Comt\'{e}}
\altaffiltext{9}{Observatoire de Besan\c{c}on, Institut Utinam, UMR 6213 CNRS, BP 1615,F-25010 Besan\c{c}on Cedex, France}
\altaffiltext{10}{Department of Astronomy, University of Michigan, Ann Arbor, MI, 48109, USA}
\altaffiltext{11}{Instituto de Astrof\'{\i}sica de Canarias, 38205 La Laguna, Tenerife, Spain}
\altaffiltext{12}{Departamento de Astrof\'{\i}sica, Universidad de La Laguna, 38206 La Laguna, Tenerife, Spain}
\altaffiltext{13}{Apache Point Observatory, P.O. Box 59, Sunspot, NM 88349-0059, USA}
\altaffiltext{14}{Department of Physics \& Astronomy, Texas Christian University, TCU Box 298840, Fort Worth, TX 76129}
\altaffiltext{15}{Department of Astronomy, University of Florida, 211 Bryant Space Science Center, Gainesville, FL, 32611-2055, USA}
\altaffiltext{16}{Astrophysics Research Institute, Liverpool John Moores University, Wirral, CH41 1LD, UK}
\altaffiltext{17}{McDonald Observatory, The University of Texas at Austin, Austin, TX, 78712, USA}
\altaffiltext{18}{Department of Physics and Astronomy, Vanderbilt University, Nashville, TN 37235, USA; Department of Physics, Fisk University, Nashville, TN, USA)}
\altaffiltext{19}{Astronomy Department, University of Washington, Box 351580, Seattle, WA 98195, USA}


\begin{abstract} 
We are carrying out a large ancillary program with the Sloan Digital Sky Survey, SDSS-III, using the fiber-fed multi-object near-infrared APOGEE spectrograph, to obtain high resolution H-band spectra of more than 1200 M dwarfs. These observations will be used to measure spectroscopic rotational velocities, radial velocities, physical stellar parameters, and variability of the target stars. Here, we describe the target selection for this survey, as well as results from the first year of scientific observations based on spectra that will be publicly available in the SDSS-III DR10 data release. As part of this paper we present radial velocities and rotational velocities of over 200 M dwarfs, with a $v \sin{i}$ precision of  $\sim$2 km s$^{-1}$ and a measurement floor at  $v \sin{i}$=4 km s$^{-1}$.  This survey significantly increases the number of M dwarfs studied for rotational velocities and radial velocity variability (at $\sim 100-200$ m s$^{-1}$), and will inform and advance the target selection for planned radial velocity and photometric searches for low mass exoplanets around M dwarfs, such as HPF, CARMENES, and TESS. Multiple epochs of radial velocity observations enable us to identify short period binaries, and AO imaging of a subset of stars enables the detection of possible stellar companions at larger separations. The high-resolution APOGEE spectra, covering the entire H band, provide the opportunity to measure physical stellar parameters such as effective temperatures and metallicities for many of these stars. At the culmination of this survey, we will have obtained multi-epoch spectra and radial velocities for over 1400 stars spanning the spectral range M0-L0, providing the largest set of near-infrared M dwarf spectra at high resolution, and more than doubling the number of known spectroscopic  v$\sin{i}$ values for M dwarfs.  Furthermore, by modeling telluric lines to correct for small instrumental radial velocity shifts, we hope to achieve a relative velocity precision floor of 50 m s$^{-1}$ for bright M dwarfs. With three or more epochs, this precision is adequate to detect substellar companions, including giant planets with short orbital periods, and flag them for higher-cadence followup. We present preliminary, and promising, results of this telluric modeling technique in this paper.

\end{abstract}


\section{Introduction}
M dwarfs are a major stellar constituent of the Galaxy and are increasingly important for exoplanet science and the quest for discovering low mass planets in or near habitable zones \citep{dressing13,kopparapu13}. High-resolution spectroscopic observations of M stars are crucial for understanding stellar astrophysics at the bottom of the main sequence and planet statistics across the HR diagram. A large, multi-epoch spectroscopic survey enables us to not only detect low-mass stellar and sub-stellar companions (including giant planets) but also to address a wide range of questions, including the statistics of stellar multiplicity, kinematics, metallicity, and activity. High-resolution spectra in the near infrared (NIR) for a range of M dwarfs will also be critical for modeling of chemical abundances and for probing the physical processes that occur in the complex atmospheres of M dwarfs. The multiplexed Sloan Digital Sky Survey (SDSS-III) APOGEE spectrograph is uniquely suited to provide many hundreds of such spectra, increasing the number of available high-resolution NIR M dwarf spectra by at least an order of magnitude.

Radial velocity (RV) surveys to detect exoplanets have mostly targeted FGK stars over the past two decades, but are beginning to explore the M dwarfs in the solar neighborhood. With temperatures below $\sim$ 4000 K, the spectral energy distributions of M dwarfs peak between 0.9 and 1.8 $\mu$m (the $Y, J, \rm{and}$\ $ H$ bands), where RV measurement techniques are less developed than in the optical. \cite{rodler2012} derived RV measurements of eight late-M dwarfs as part of a large survey by  \cite{deshpande12} to monitor RV variability among late-M dwarfs (M5.0 - M9.5) using the NIRSPEC spectrograph \citep{mclean98}. At a spectral resolution $R \sim 20000$, in the $J$ band, they obtained RV precisions between 180 m s$^{-1}$ and 300~m~s$^{-1}$. \cite{blake2010} targeted a magnitude-limited sample of M dwarfs and ultracool dwarfs using NIRSPEC over a period of six years. Employing a forward-modeling technique in the $K$ band \citep{blake07} at $R \sim 25000$, they report a RV precision of 50 m s$^{-1}$ for a bright, slowly rotating M dwarf and 200 m s$^{-1}$ for slowly rotating, ultracool dwarfs. \cite{bean2010} used ammonia gas cell as an iodine analog in the NIR on the very high ($R\sim 100000$) spectrograph CRIRES \citep{kaufl04} in the $K$ band and achieved an RV precision of $\sim$ 3--5 m s$^{-1}$ over long timescales. 

Though this level of precision is a big step forward, it is restricted to wavelength regions overlapping the ammonia cell and for relatively bright stars. In addition, the process also requires very high spectral resolution, which is not generally available. New infrared instruments like Habitable Zone Planet Finder \citep{mahadevan12} and CARMENES (\citealt{quirrenbach10}) are now being designed to try to detect Earth-mass planets in the habitable zones of nearby low-mass  M dwarfs. Target selection for these surveys (which tend to be very expensive in terms of telescope time) will benefit immensely from the characterization of activity and binarity in a large sample of M dwarfs.

Fully convective, late-type M dwarfs show strong magnetic activity (\citealt{morin08}; \citealt{browning08}; \citealt{reiners2008}, \citealt{reiners2009}, \citealt{reiners2010}). A relation between stellar activity and projected rotational velocity ($v \sin i$) is well established for FGK stars  \citep{noyes84} and appears to govern the M dwarfs as well. Therefore, $v \sin i$ acts as a good proxy for activity. In general, fast rotating M dwarfs are more active than their slow rotating counterparts. \cite{reiners2008} showed that early M dwarfs are slow rotators while stars M5 and later, a boundary where they become fully convective (Chabrier $\&$ Baraffe, 1997), are generally fast rotators ($v\,\sin{i}$ $>$ 10 km s$^{-1}$). A study of a sample of 27 M dwarfs over a period of 6 years by \cite{gomes12} found RV variations with a amplitudes $\ge$5 m s$^{-1}$ among 36$\%$ of their stars, which they attributed to magnetic cycles. Therefore, measuring $v \sin i$ to identify slowly rotating stars that are likely less active (or at least likely to exhibit less RV noise) is an important component of any search for low-mass companions to low-mass stars.
 
RV surveys of solar-type stars have estimated the frequency of giant planets at $\sim$7$\%$ for FGK stars \citep{marcy00b, udry07} within 10 AU. The frequency of brown dwarf (BD) companions in RV surveys, however, falls down to about 0.6-1$\%$ at similar orbital distances \citep{marcy00a, grether06}. When restricting the sample to "hot Jupiters" (at a~$<$~0.1 AU), the frequency of giant planets is $\sim$1.2$\%$, and 2.5$\%$ for a~$<$~1AU. The expected frequency of close-in giant planets (a~$<$~1AU) from M dwarf survey at the Hobby-Eberly Telescope (HET) \citep{endl03} is roughly $<$1.3$\%$ \citep{endl06}.

On the other hand, \citet{metchev09} found a frequency of 3.2$\%$ of BDs from adaptive optics survey of 266 young solar type stars at larger distances between 28 - 1590 AU, and \cite{lafreniere07} found a frequency of 1.9$\%$ of BDs companion at $\sim$ 25-250 AU. \cite{jodar13} have found that the volume-limited binary fraction of early-to-mid type M dwarfs of 8.8$\%$ from a sample of 451 M dwarfs at less than 25 pc (at orbital distances less than 80 AU). When including literature data at larger orbital distances these authors find a binary fraction of $\sim$20.3$\%$.

Exploring these large orbital distances is probably beyond the capabilities of the APOGEE M Dwarf Survey, but we will be able to discover close-in BDs around M dwarfs, and therefore place constraints on the frequency of close-in BDs and maybe giant planets around M-dwarfs.

The well-established correlation between metallicity and planet occurrence seems to also apply to M dwarfs \citep{neves13, rojas-ayala12, terrien12a} The different between mean 
metallicity of the M-dwarf stars with and without giant planets is ~0.20 dex \citep{neves13}, from a sample of 102 M dwarfs in the HARPS RV survey \citep{bonfils13}.

Imaging and RV observations of stellar multiple systems enable direct measurements of the physical properties of stars and provide a key window into the process of star formation. Comprehensive surveys probing a wide range of orbital separations using multiple observational techniques have been carried out for F, G, and K stars (\citealt{duquennoy1991,raghavan2010}), but sample sizes for similar surveys targeting low-mass stars remain small (\citealt{fischer1992,marchal2003,delfosse2004}). There are, however, clear indications that the overall rate of occurrence of multiple systems is a strong function of stellar mass \citep{lada2006, raghavan2010}. \cite{monet13} find that $\sim$~6.5$\%$ of all M dwarfs stars host a giant planet with 1 $<$ M$_{J}$ $<$ 13 M$_{J}$ and a $<$ 20 AU. 

The distribution of orbital separations and mass ratios in multiple systems can be compared to numerical simulations of star formation. Individual binary star systems that eclipse, or that can be spatially resolved, can be used to make precise measurements of fundamental physical properties of stars, such as mass and radius (see \citealt{southworth2011} and \citealt{torres2012} for recent examples). These measurements provide primary observational constraints on theoretical models of stellar structure and evolution. At the bottom of the stellar main sequence only a handful of these systems are known, and measurements deviate from theoretical expectations at the level of a few percent\citep{chabrier2007,torres2012b,feiden12,terrien12b}. For this reason, it is particularly important to identify bright low-mass binaries over a wide range of orbital periods. 

Previous large spectroscopic surveys designed to study M dwarf stellar properties \citep[e.g.][]{reiners12, jenkins09} have included $\sim$ 300 M dwarfs. In this paper, we describe an ongoing, extensive M dwarf spectroscopic survey of 1404 M dwarfs designed to detect low mass companions, quantify the statistics of stellar multiplicity, and measure basic stellar physical parameters at the bottom of the main sequence. Using 253 of the 1404 stars that have been observed so far, we illustrate the techniques employed to measure observational and model-derived stellar parameters, and we demonstrate the concurrent use of RV monitoring and AO imaging to search for stellar companions to our sample of M dwarfs. The reduced and calibrated spectra for this first set of stars will be publicly available as part of the SDSS-III DR10 data release in Summer 2013 and spectra for all stars will be released in Dec 2014 as part of the SDSS-III final data release.  In addition to providing spectroscopic rotational velocities and radial velocities, this paper serves as an introduction to this unparalleled data-set, lays out our target selection choices, and highlights some of the science investigations we are undertaking. 

\section{The APOGEE M Dwarf Survey}

The Apache Point Observatory Galactic Evolution Experiment \citep[APOGEE,][]{majweski10} is a high-resolution (R$\sim$ 22500), near-infrared ($H$ band), multi-object, fiber-fed, and cryogenically cooled spectrograph \citep{wilson10,wilson12}. The instrument is part of SDSS-III \citep{eisenstein11}, attached to the 2.5m SDSS telescope \citep{gunn06} at Apache Point Observatory, and covers a wide field of view ($3^{\circ}$ diameter). The instrument can observe up to 300 targets simultaneously on a three-segment mosaic of Teledyne H2RG 2048 x 2048 detector arrays. Each detector has a wavelength range of $\sim$ 0.07 $\mu$m and covers 1.514 - 1.581 $\mu$m (blue), 1.586 - 1.643 $\mu$m (green), and 1.643 - 1.696 $\mu$m (red), respectively. The entire assembly is enclosed in a vacuum shell and is intrinsically very stable.

Our M dwarf survey is an ancillary program to the main SDSS-III APOGEE survey, which began in September 2011 and will end in June 2014. The observations are made during bright time. The light from the telescope is focused onto standard SDSS plugplates which have holes drilled in them to accommodate fibers. A total of 300, 120 $\mu$m fibers are attached to the plugplate. Each fiber has a 2" diameter. The 300 fibers on each plate are allocated as follows: 230 are placed on science targets, 35 on telluric standards (hot stars used to calibrate and remove telluric features), and another 35 fibers are placed in sky regions devoid of celestial objects in order to obtain sky spectra (Nidever et al. 2013). Multiple visits are made to many of the APOGEE fields during the course of the survey, allowing us to monitor RV variability among our sample of stars. 

The main goal of APOGEE is to measure radial velocities and chemical abundances of 10$^5$ red giant branch stars spanning various Galactic environments such as the bulge, disk, bar, and halos (Majewski et al, in prep). Hence, the SDSS-III survey footprint primarily spans these regions. A detailed discussion of the APOGEE target selection and field plan can be found in Zasowski et al. (2013). Targets from our survey were generally distributed among fields with 3--24 visits, though some M dwarf targets are also present on single-visit fields. The M dwarfs can be identified in the DR10 release data by looking for bit flag 19. Spectra with this bit flag correspond to this program.

The APOGEE data reduction process is described in detail in Nidever et al. (2013). Products from the data reduction pipelines include: (1) apCframe files containing individual 1-D, dithered, wavelength calibrated spectra that represent individual exposures taken over the course of a single visit; (2) apVisit files consisting of co-added apCframe spectra from a single visit that are wavelength calibrated, flux normalized, and telluric corrected; and (3) apStar files including RV-shifted, co-added apVisit spectra and a best matching synthetic spectrum. A detailed list of various file formats are given in Nidever et al. (2013) and also described in detail as part of the SDSS-III DR10 release. 

The goal of this survey is to characterize M dwarfs through measurement of projected rotational velocities, absolute RV, and metallicity, and to discover new low mass companions. This multi-epoch spectroscopic survey is sensitive enough to detect low-mass stellar and sub-stellar companions to M dwarfs and will also shed light on the statistics of stellar multiplicity at the bottom of the main sequence. Furthermore, these observations will create an atlas of high-resolution spectra in the NIR covering a wide range of sub-spectral types. These spectra are critical for modeling chemical abundances and probing the physical processes that occur in the complex atmospheres of M dwarfs. As an ancillary project to the APOGEE survey, this project has been awarded 6000 fiber hours \footnote[1]{One fiber hour is an hour long exposure of a single target.} over the SDSS-III survey.

\subsection{Target Selection}

Targets for the APOGEE M dwarf survey primarily consist of stars from the all-sky Le\'pine and Shara proper motion North catalog \citep[][LSPM-N hereafter]{lepine05} and the all-sky catalog of bright M dwarfs \citep[][LG11 hereafter]{lepine11}. 

The LSPM-N catalog contains stars with proper motions $\mu$ $>$ 150~mas~yr$^{-1}$. We apply magnitude and color cuts to select red stars that fall within the APOGEE target field plan ($7<H<12$; $V-K>5.0$; $0.4<J-H<0.65$; $0.1<H-K_{\rm{s}}<0.42$). Contamination from M giants in a sample of M dwarfs is not uncommon, but they can be distinguished from dwarfs using infrared color cuts. Stars that pass these selection criteria and lie on planned APOGEE main survey fields are selected as targets.  Aspects of this selection are also discussed in Zasowski et al. (2013),  which describes target selection for the entire APOGEE survey, but are presented in more detail here.

\begin{figure}[t!]
\includegraphics[scale=0.5]{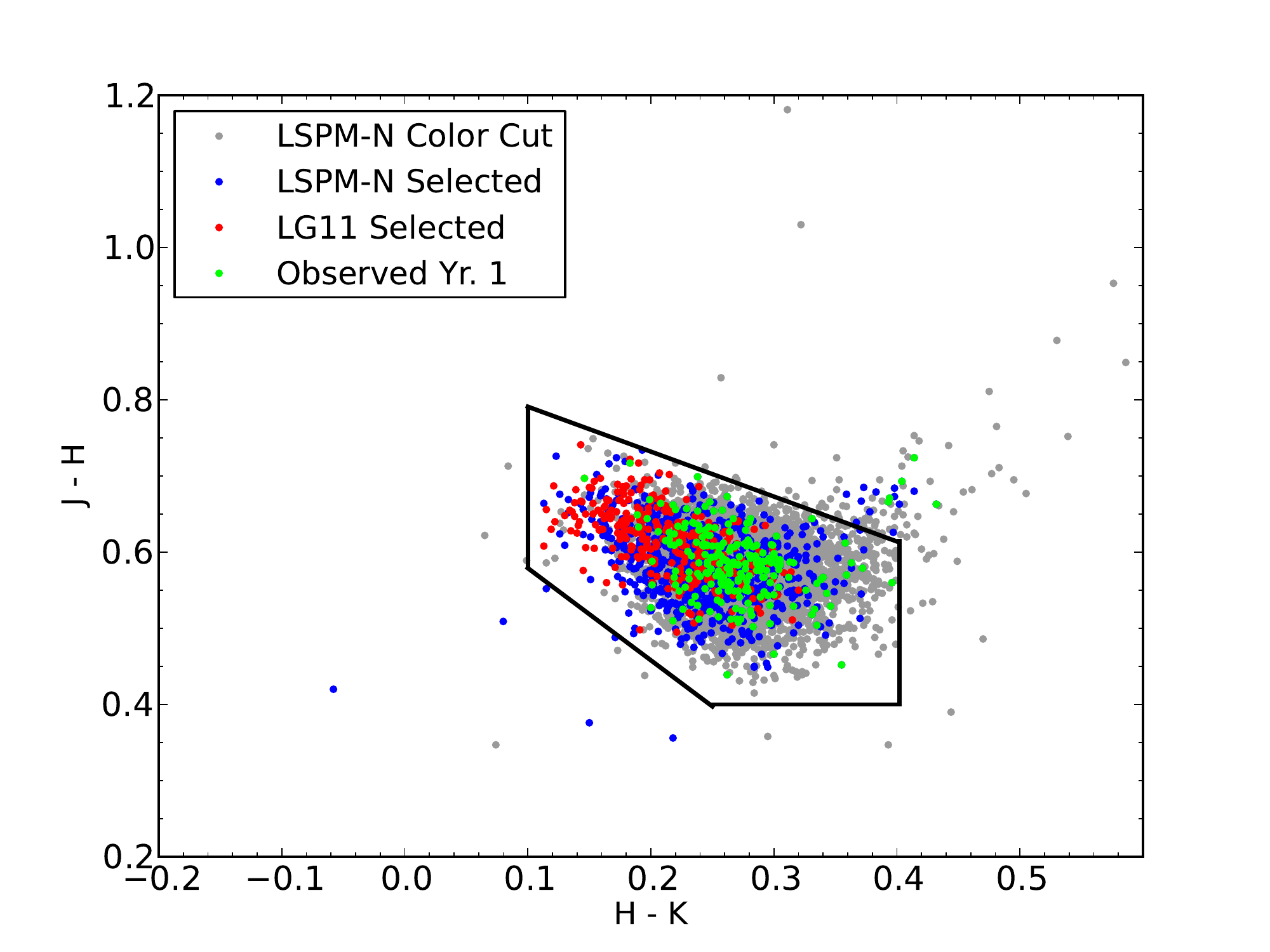}
\caption{Infrared color-color diagram of our sample. The gray points are the 7059 LSPM-N stars obtained using the magnitude and color-cuts described in Section 2.1. The LSPM-N (blue), LG11(red), and stars observed in year one (green) are also plotted. Stars within the box, defined as the ``red dwarf box'' in LG11, are likely to be dwarfs, though some contamination by giants is possible. \label{cmd_giants_dwarfs}}
\end{figure}

The LG11 catalog became available only after we had submitted targets for initial plate drilling. LG11 is an end result of a careful application of infrared color and magnitude cuts to stars in the SUPERBLINK proper motion survey with the goal of selecting bright M dwarfs.  Stars in the catalog are limited by apparent magnitude $J <$ 10 and $\mu > $ 40~mas~yr$^{-1}$. When the catalog became available we used it to select targets for future plate drillings. To comply with our observational constraints, we apply declination ($\delta > 0^{\circ}$) and magnitude cuts to their sample ($7<H$ to avoid saturating the APOGEE detector array). An important point to emphasize is that while our LSPM-N cuts were designed to select stars of spectral type M4 and later, our LG11 selection includes M dwarfs of all spectral types. The target selection in color-color space is illustrated in Figure \ref{cmd_giants_dwarfs}. The region defined by equations 9-13 from LG11 should contain stars that are potentially red dwarfs, while those objects outside the box, especially the reddest objects, are more likely M giants. We find that the majority of our target stars selected from LSPM-N (blue points) and LG11 (red points) lie within this box in color-color space.   

\begin{figure}[t!]
\includegraphics[scale=0.5]{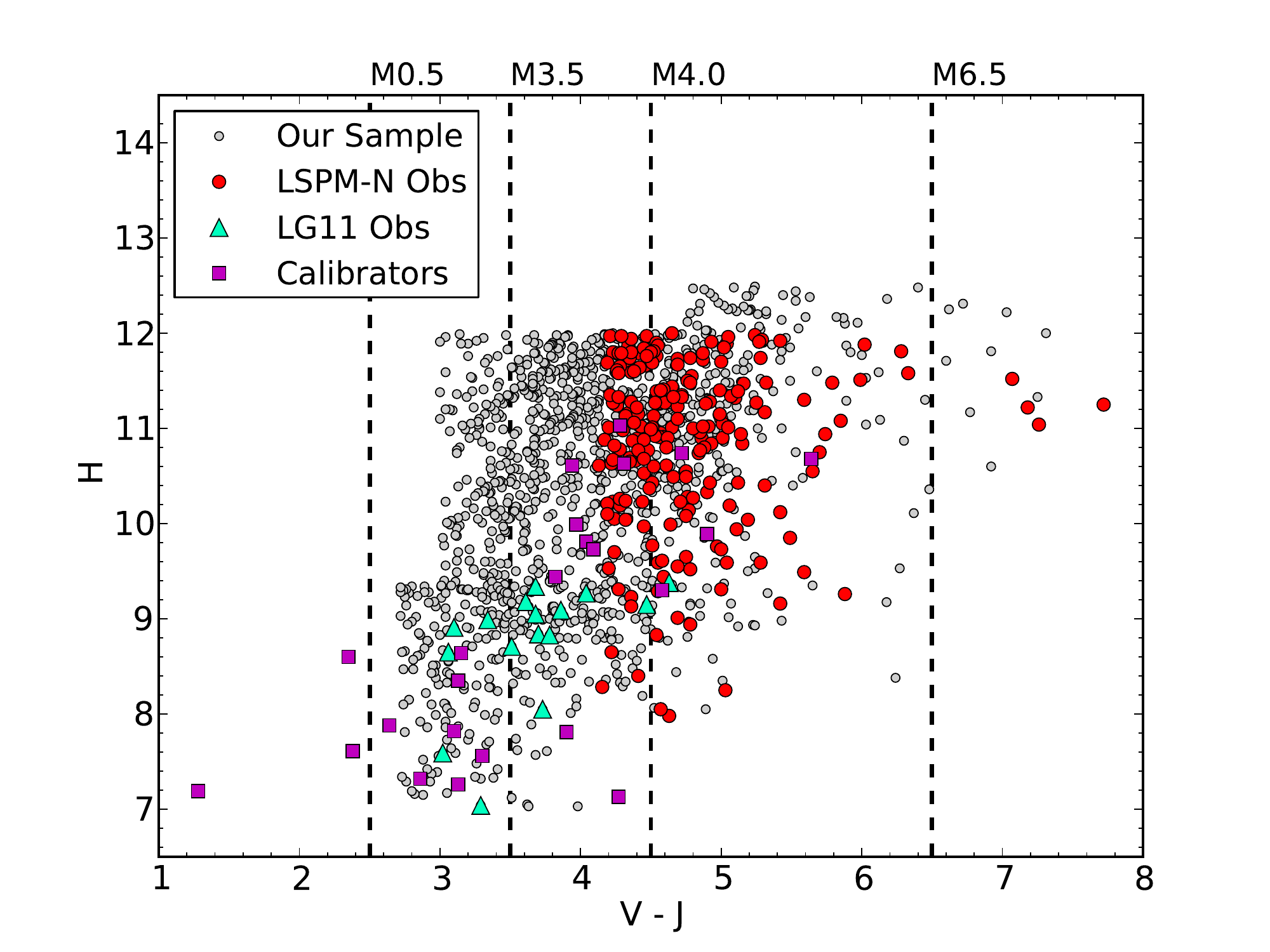}
\caption{The entire APOGEE M dwarf sample (gray filled circles) and calibrators (magenta squares). The stars observed in the first year (Sept. 2011 - July 2012) are shown in color: LSPM-N (red filled circles), LG11 (green filled triangles). The vertical dash lines mark approximate positions of M dwarf spectral sub-types.\label{all_mdwarfs}}
\end{figure}

Each APOGEE field consists of a circular plate with a radius of 1.49$^{\circ}$. The plate has a bolt at the center with a radius of 5$\arcmin$ that cannot accept fibers. We assembled our final target list by cross-correlating our master target list with the area covered by each APOGEE plate. We anticipate observations of more than 1400 M dwarfs over the course of the survey. 

\begin{figure}
\includegraphics[scale=0.5]{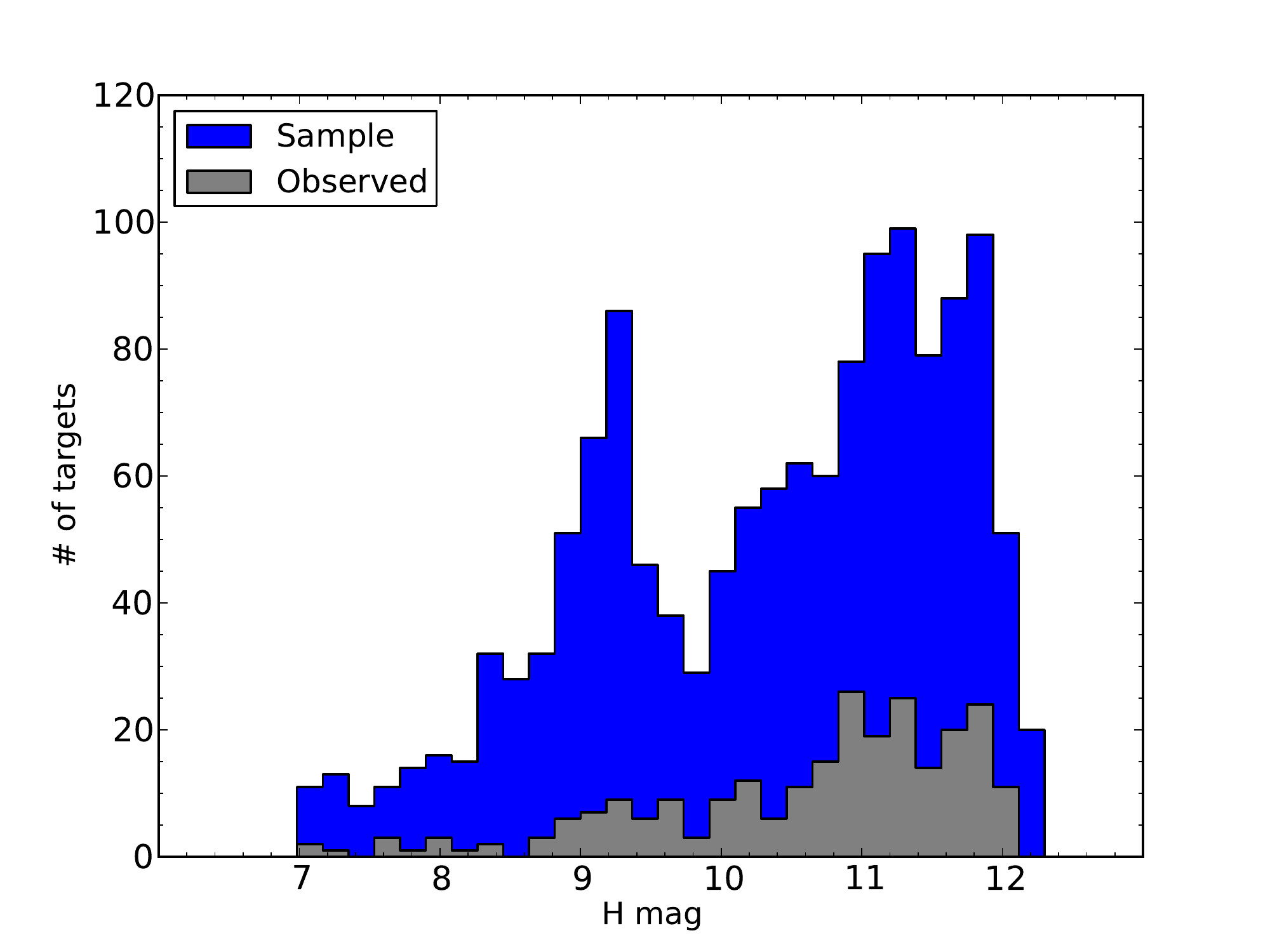}
\caption{Distribution of $H$ magnitudes of the targets observed during the first year (gray; Sept. 2011 - July 2012)~and our entire sample (blue).\label{hmag}}
\end{figure}

In the first year of the SDSS-III APOGEE survey 285 of these stars were submitted during plate drilling, of which 253 were observed (the remaining 32 were lost due to fiber conflicts with other targets or fell off the inner or outer plate edge when plate centers were slightly adjusted during final design). Figure \ref{all_mdwarfs} shows our entire sample in terms of color, magnitude, and approximate spectral types. The LSPM-N and LG11 (gray) is our total sample. The 253 targets observed during the first year are plotted as red (LSPM-N) and blue points (LG11).  As a result of the timing of the LG11 catalog release, most of stars observed during the first year were selected from LSPM-N and have spectral type of M3.5 and later. The observations of these stars have produced 1127 spectra. The distribution of their $H$ magnitude is shown in Figure \ref{hmag}. 

We have also deliberately targeted some calibration stars, which include RV standards from the California Planet Survey, $v \sin i$ standards from literature \citep{jenkins09}, stars from the MEarth Project \citep{nutzman08} and M dwarfs in the \emph{Kepler} field that are known to be active \citep{ciardi11,walkowicz11}. The selection of calibrators do not follow the stringent magnitude and color-cuts used as listed above, with the exception of bright magnitude cut: $H>7.0$. Figure \ref{all_mdwarfs} shows the sample (magenta squares). These stars are listed in Table \ref{table2}.  In total, 25 calibration stars have been observed in the first year.

In order to explore the relationship between M dwarf subtype and features in the APOGEE high-resolution spectra, we constructed a set of spectral templates, using the combined spectra in the apStar files. For each subtype, we grouped the spectra by their V-J color, which we calibrated to spectral type using the relation from \cite{lepine11}. and bins centered on integer spectral types (e.g. M3 contains all stars with V-J colors corresponding to M2.5 to M3.5).  It is important to note that the V magnitudes for many of these targets are based on scans of photographic plates and are therefore only reliable to $\pm0.5$ mag. This error in V corresponds to approximately 1-2 subtypes. We then shifted these spectra to a common RV, interpolated to a common wavelength grid, and filtered each spectrum for strong remaining sky lines and other artifacts. With these cleaned spectra, we constructed an unweighted average spectrum for each spectral type. The spectral type bins contained as few as three (M1) or four (M8) spectra and as many as 95 (M4) spectra. Finally, we constructed artificial spectra for each 0.1 subclass by linearly interpolating between subclasses for each pixel.

\begin{figure}[htb!]
\minipage{0.55\textwidth}
   \includegraphics[width=\linewidth]{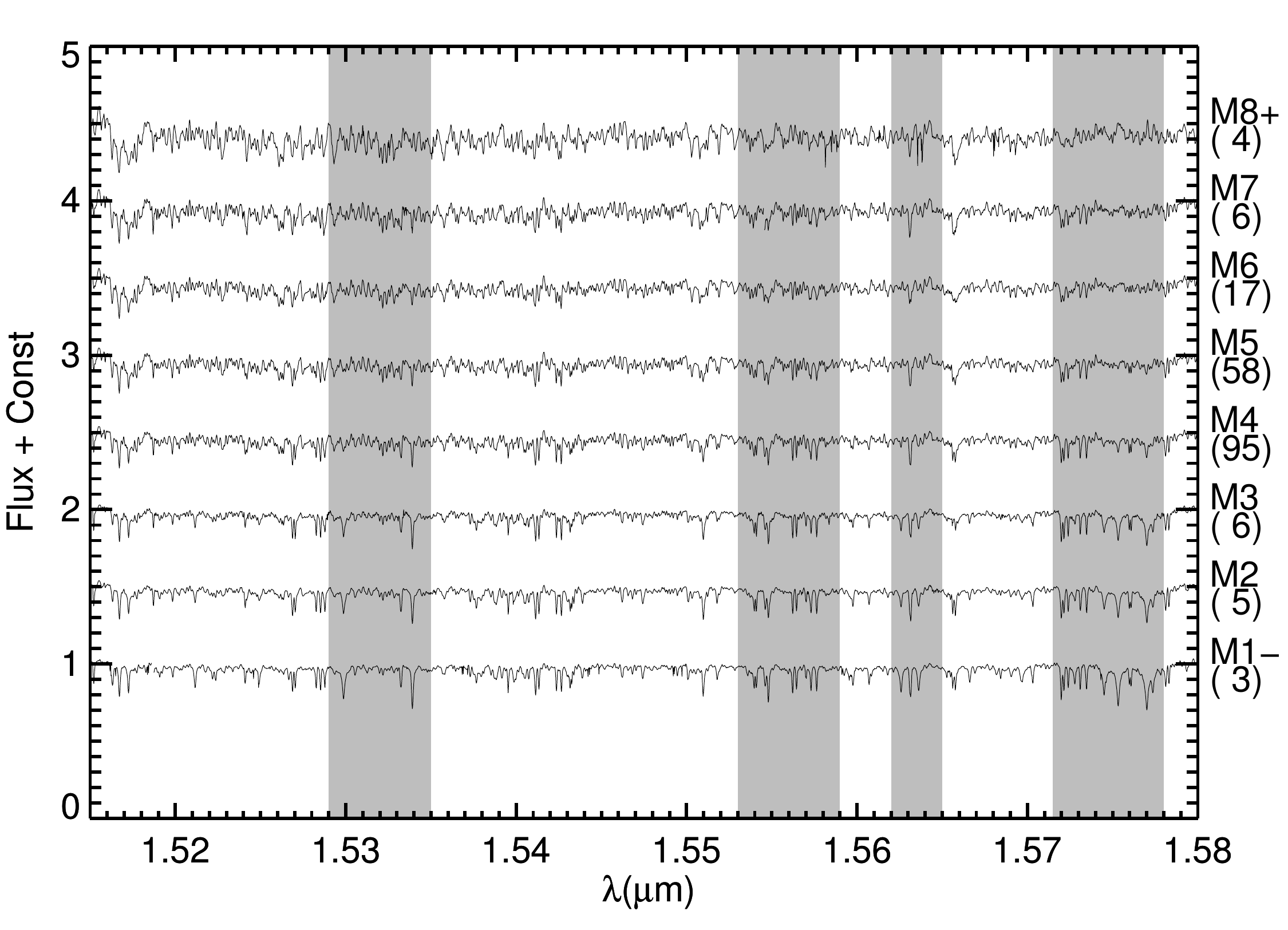}
 \endminipage\vfill
\minipage{0.55\textwidth}
   \includegraphics[width=\linewidth]{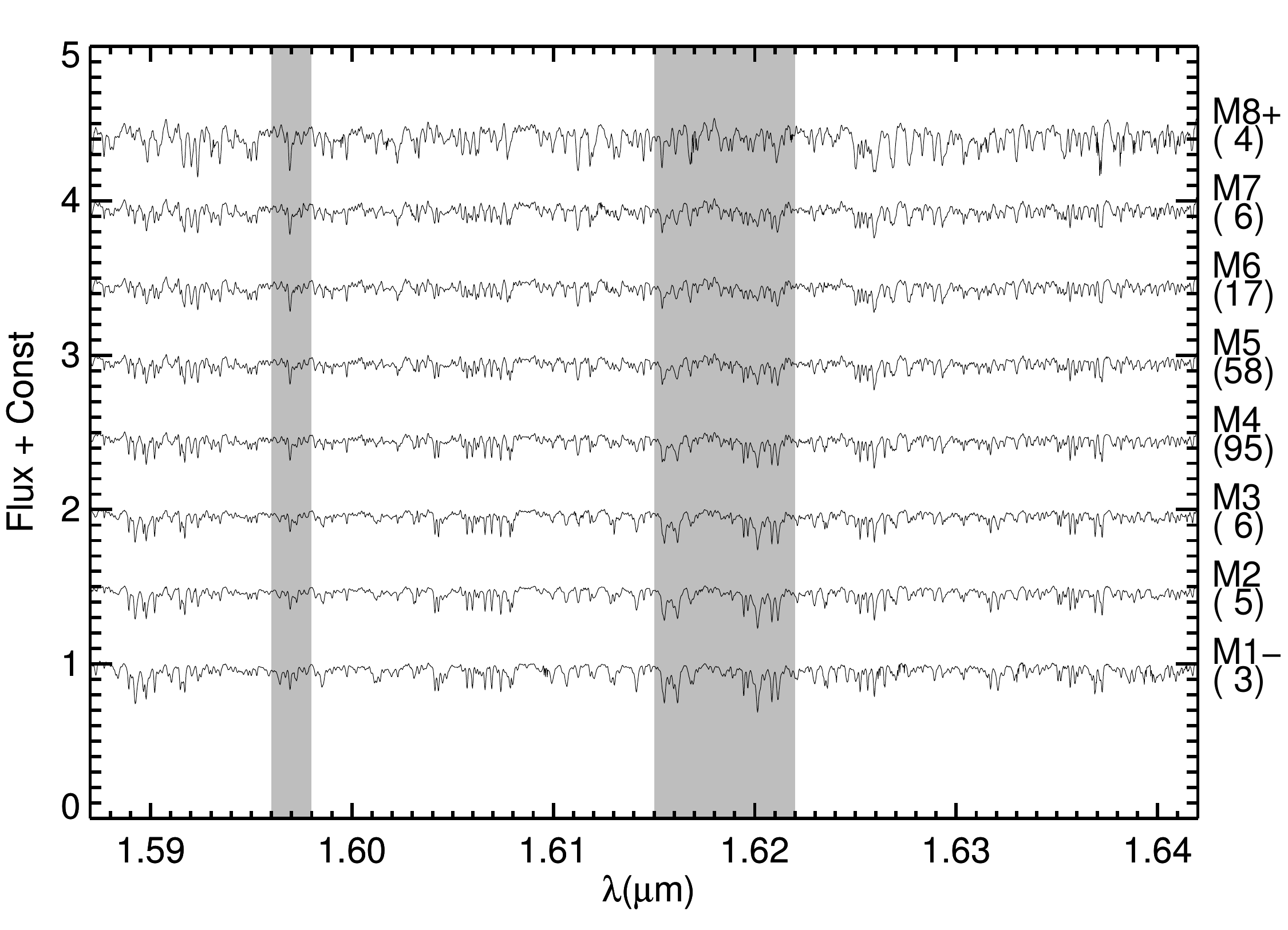}
 \endminipage\vfill
 \minipage{0.55\textwidth}
   \includegraphics[width=\linewidth]{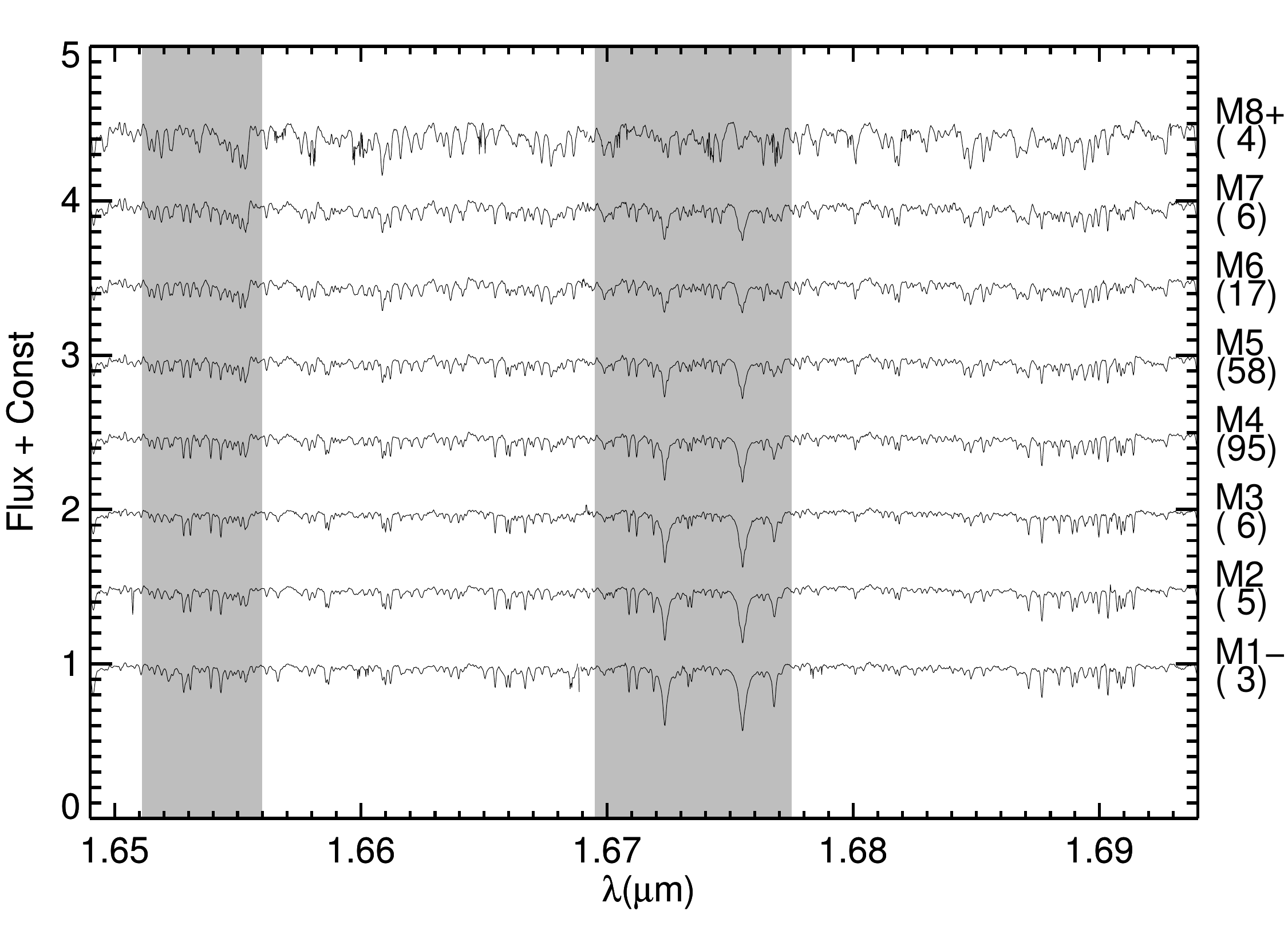}
 \endminipage\vfill
\caption{The average spectrum for each approximate spectral type, binned by V-J color. The gray regions highlight the spectral-type sensitive regions that were used for spectral type estimation, and the numbers in parentheses indicate the number of targets in each bin.The three plots correspond to blue chip (top), green chip (center), and red chip (bottom). \label{sptype}}
\end{figure}

We then constructed a routine to apply these templates to estimate the spectral type for any APOGEE M dwarf spectrum. We selected regions that showed the most sensitivity to spectral type (Figure \ref{sptype}), and set the remaining spectrum to a flat continuum. For each target we then performed a cross-correlation on these filtered spectra with the equivalently-filtered template spectra. The 0.1 subclass with the highest cross-correlation value was taken to be the estimated spectral type. Although this method is imprecise, it can be efficiently applied to estimate spectral types for large sets of targets with poorly constrained visual magnitudes. And despite the high uncertainties in the V-J colors, the likely range of $v\sin i$ values for our targets, and the possible multiplicity of many targets, these templates clearly demonstrate the regions of M dwarf spectra in the H-band that are the most sensitive to spectral type.

In addition to the M dwarfs directly selected as part of our ancillary program other M dwarfs are sometimes serendipitously observed by APOGEE. We do not discuss those in this intermediate data release paper, but they will be included in the final survey analysis. 
\newline
\section{Initial Results}
Our APOGEE M dwarf spectra span a wide range of signal-to-noise (S/N). Depending on the target brightness and observing conditions, the S/N of individual APOGEE visits varies from a few tens to several hundred per resolution element. The number of visits available for each target for the stars observed in the first year of the survey also varies: some targets only have a single visit so far, while others have 12 or more. In the following subsections we present initial results based on the analysis of spectra from Year 1 of the survey.

\subsection{Projected Rotational Velocities: $v \sin i$}

We measured projected rotational velocities for all of our targets using cross-correlation techniques. To maximize the precision of our measurement, we use the entire APOGEE spectral range. The first method cross-correlates a rotationally broadened synthetic spectrum against an observed target spectrum, and derives $v \sin i$ by maximizing the amplitude of the resulting correlation peak. The second method cross-correlates the observed spectrum of a slowly rotating template star against the observed target spectrum and measures the full-width at half-maximum (FWHM) of the resulting correlation peak.  This width is compared to tabulated values obtained by rotationally broadening the template spectrum and cross-correlating it against the unbroadened version of itself.  Both of these methods are described in further detail below.

\subsubsection{Projected Rotational Velocities: Method I}
Our first method for measuring the $v \sin i$ of a target star uses a synthetic template spectrum with stellar characteristics ($T_{\rm{eff}}$, $\log{g}$, and $[M/H]$), spectral resolution, and wavelength sampling that match those of the target.  The template is rotationally broadened over a wide range of $v \sin i$ to generate a suite of broadened synthetic templates. Each broadened template is cross-correlated against the target spectrum, and the amplitude of the resulting correlation peak is measured.  The correlation amplitude varies slowly with the template $v \sin i$, and the peak of this function determines the best $v \sin i$ estimate. This technique utilizes the entire free-spectral range that we deem to be free of telluric absorption and night sky OH emission, and we have used it in several past analyses of cool star spectra \citep[e.g,][]{vaneyken12,bender08}.  Observations with S/N or spectral resolution insufficient for more traditional approaches that examine the line profiles of specific individual spectral features \citep{gray_phot}, or where the spectral line density implies significant line blending, can still utilize this correlation based approach.

\begin{figure}
\includegraphics[scale=0.50]{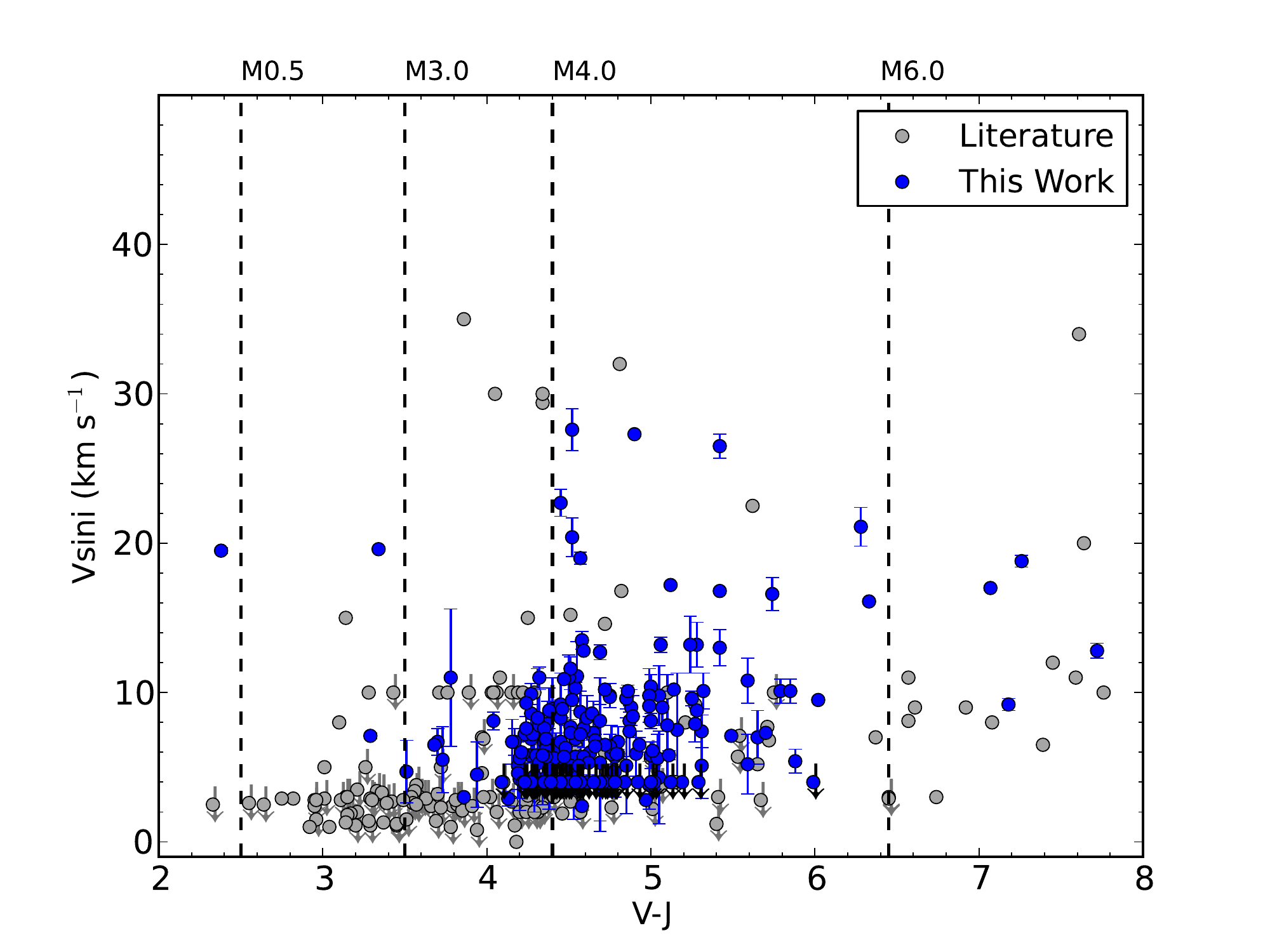}
\caption{Measurements of $v \sin i$ for M dwarfs, including the measurements presented here (blue filled circles) and those from the literature (gray filled circles), for a range of V-J colors. The dashed lines mark the approximate positions of M dwarf spectral sub-types.  \label{vsini_plot}}
\end{figure}

Carrying out this analysis requires \textit{a priori} knowledge of the target star's $T_{\rm{eff}}$, $\log{g}$, and $[M/H]$, so an appropriate synthetic template can be selected. An inaccurate choice at this point can introduce systematic errors in the derived rotational velocities. In the initial APOGEE data reduction pipeline, each apVisit spectrum is cross-correlated against a large grid of synthetic spectra as part of an initial spectroscopic characterization. The section of this grid that corresponds to $T_{\rm{eff}}$ $<$ 4000 K, and relevant for our M dwarf sample, is composed of synthetic spectra from the standard BT-Settl grid.  These initial correlation-based estimates return $T_{\rm{eff}}$ to the nearest 100 K, and $\log{g}$ and $[M/H]$ to the nearest 0.5 dex.  Most APOGEE spectra are subsequently passed through the APOGEE spectra analysis pipeline (ASPCAP), which re-derives these parameters with much finer precision.  However, ASPCAP results are currently not regarded as reliable for cool stars with $T_{\rm{eff}}$ $<$ 4000 K.  As such, we average the $T_{\rm{eff}}$, $\log{g}$, and $[M/H]$ derived for each visit spectrum from the synthetic grid, and use these values to select the appropriate BT-Settl model \citep{allard97} for our $v \sin{i}$ analysis.

\begin{figure}[t!]
\includegraphics[scale=0.5]{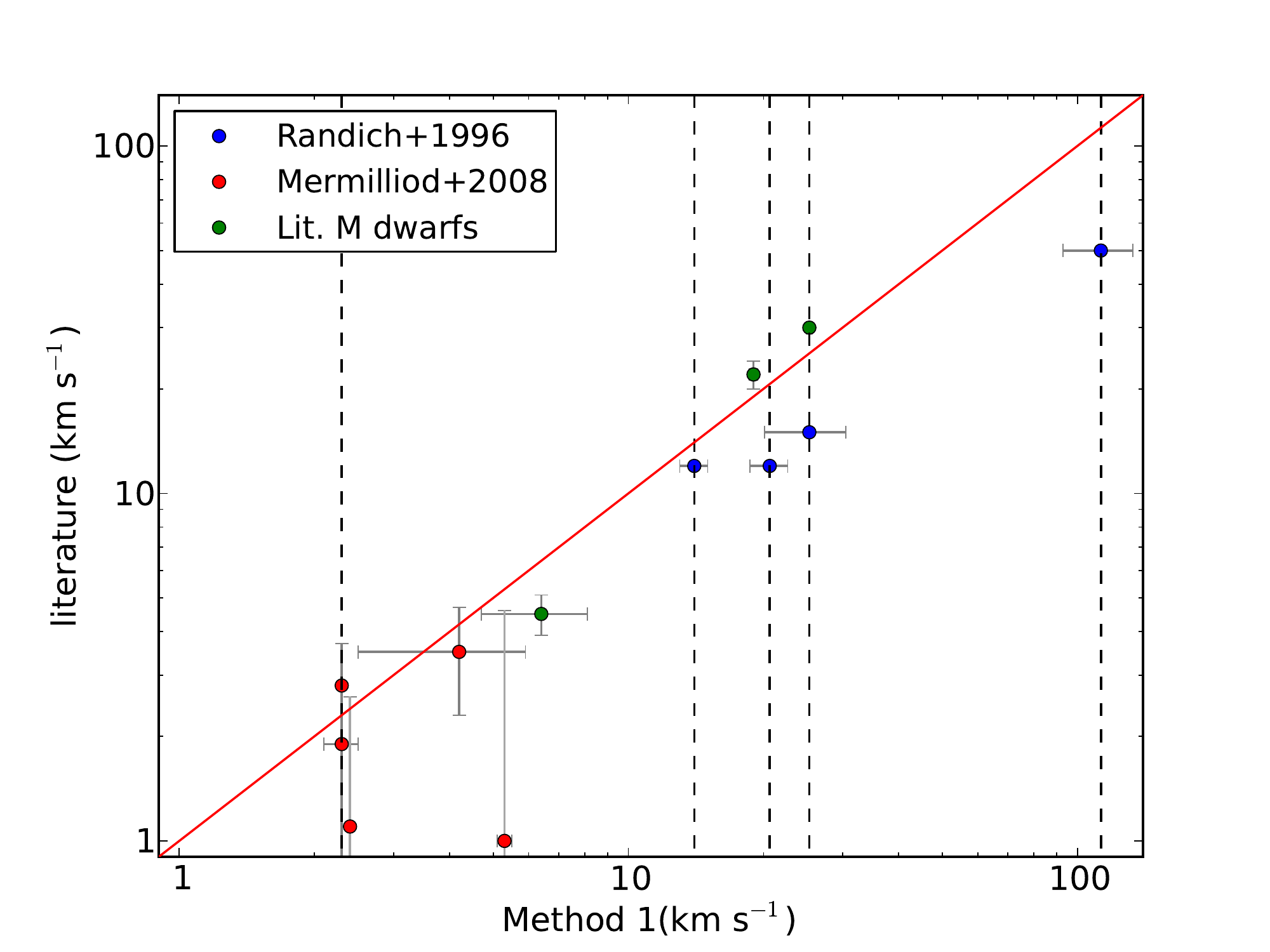}
\caption{Comparison of literature $v \sin i$ measurements of stars with Method 1. Stars in this figure include spectral types of A5 to M6. Literature M dwarfs are listed in Table \ref{table2}. The solid line indicates 1:1. Literature stars that do not have reported errors are marked by dash lines. \label{vsini_lit_comp}}
\end{figure}

The spectral resolution of a model must be degraded to match the resolution of the corresponding target spectrum. APOGEE's spectral resolution varies across the spectral range, and from fiber to fiber \citep{wilson12} across the spectral range, but the Line Spread Function (LSF) for each fiber is automatically parametrized by the reduction pipeline as a Gauss-Hermite polynomial and stored along with the individual visit spectra. APOGEE targets with multiple visits are not guaranteed to be carried out using the same spectrograph fiber, although in practice they often are. Targets with visits distributed over multiple fibers will result in visit spectra with slightly different LSFs; these effectively get averaged together when the data pipeline co-adds the individual visit spectra. For simplicity, we only consider a target's average LSF, and approximate it as a Gaussian to derive the corresponding effective spectral resolution.  We then degrade the synthetic template spectrum to the derived resolution.

We rotationally broaden the synthetic templates using a four parameter non-linear limb-darkening model \citep{claret12,claret00, gray_phot}, with parameters appropriate for the characteristics of the stellar spectrum and the APOGEE H-band bandpass, over a range of v$\sin{i}$ from 3 km s$^{-1}$ -- 100 km s$^{-1}$.  Finally, we resample both the APOGEE spectra and synthetic models to log-lambda wavelength space \citep{tonry79} in preparation for cross-correlation. The apStar files produced by the APOGEE data pipeline contain co-added spectra for each multi-visit target using two different co-adding schemes. Our $v \sin i$ analysis measures both co-added apStar spectra, and also each of the individual apVisit spectra. The $v \sin i$ from the individual visit spectra are then combined using a weighted average, with the weights derived from the S/N of each visit.  We take the standard deviation of the distribution as the measurement precision on $v \sin i$, and impose the following additional rules: (1) single visit spectra have default precision of 2 km s$^{-1}$; (2) multi-visit spectra with three or fewer visits have a minimum precision of 1 km s$^{-1}$.  In addition, we set a conservative floor in our ability to measure $v \sin i$ at 4 km s$^{-1}$, which also corresponds to the minimum $v \sin i$ where the broadening kernel is resolved at APOGEE resolution and sampling.

We further tested the method by measuring $v \sin i$ of main sequence stars in the main APOGEE survey. We found 9 stars with spectral types of A5 -- K0 with a large range of previously reported $v \sin i$ values \citep{randich96,mermilliod08}. Figure \ref{vsini_lit_comp} illustrates this effort. Most of our stars are within 1-$\sigma$ of the 1:1 line.

\subsubsection{Projected Rotational Velocities: Method II}

Our second method measures $v \sin i$ by cross-correlating object spectra against a slowly rotating template spectrum of similar spectral type and measuring the FWHM of the cross-correlation peak. This process is well established in literature \citep[e.g.][]{bailer-jones04} and assumes that the line profile is primarily dominated by rotation. 

Four slowly rotating stars are used as templates: 2M05470907$-$0512106 with $v \sin i=~$4.5\,km\,s$^{-1}$ \citep{jenkins09} and 2M19125504$+$423937, 2M19121128$+$4316106, and 2M19332454$+$4515045 with rotation periods of 48.5, 19.0, and 42.6 days, respectively, determined from Kepler photometry \citep{McQuillan:2013wb}. We used the apVisit and apStar files in the analysis.  

\begin{figure}[t!]
\includegraphics[scale=0.5]{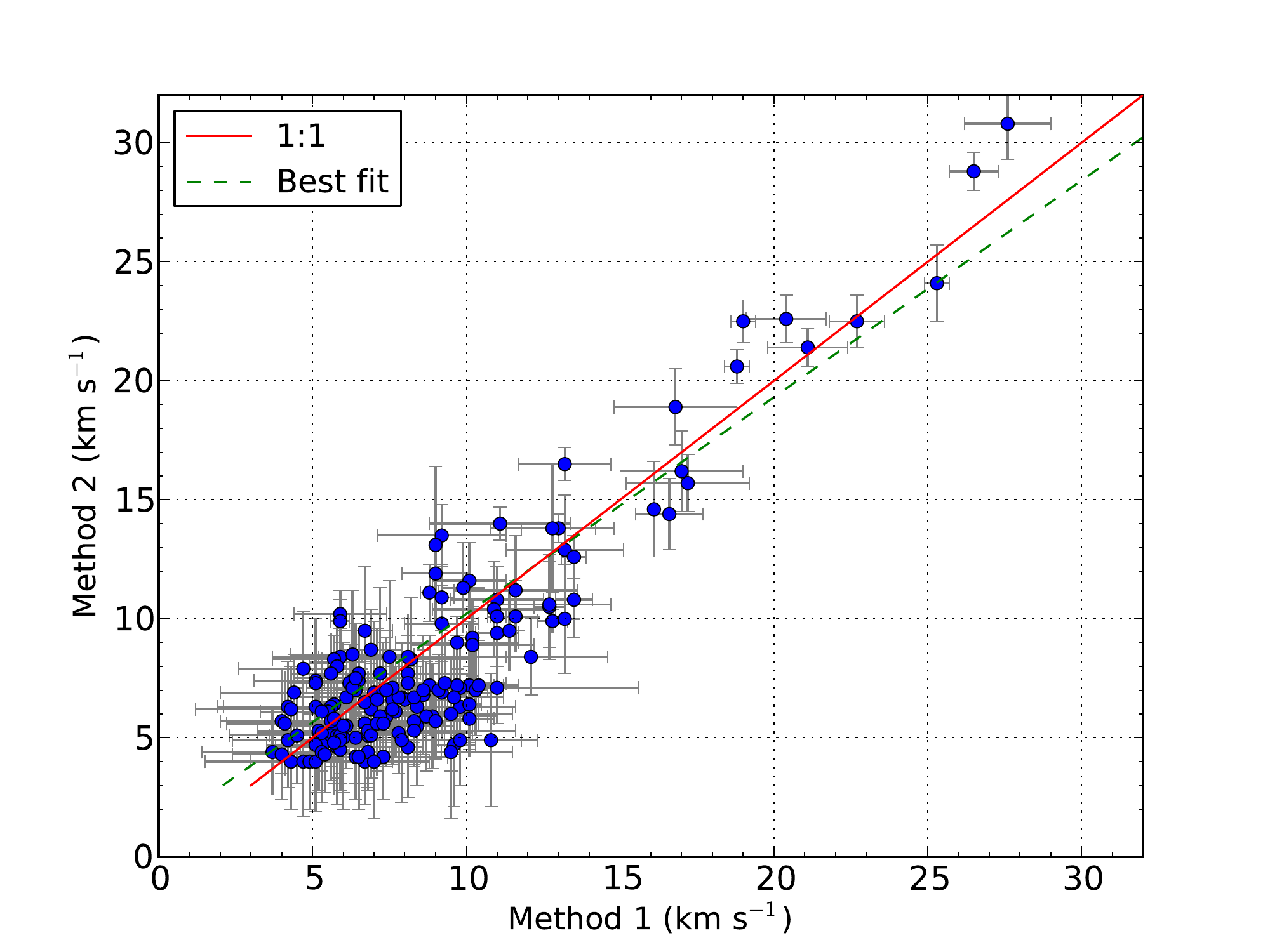}
\caption{Comparison of $v \sin i$ measurements using Method 1 and Method 2, as described in Section 3.1. The solid line represents perfect agreement. The linear regression fit (y = Bx + A; B = 1.099, A = -1.212) is indicated by a dash green line.  \label{vsini_comp}}
\end{figure}

Prior to cross-correlating, the spectra are first continuum normalized and strong telluric absorption lines and OH emission lines are masked out. In the H-band, the number of potential contaminating sky emission lines can be large, and even after sky subtraction the flux in the residual sky emission lines sometimes exceed the stellar flux by a considerable amount.  If not masked, these telluric features can dominate the cross-correlation signal. To limit the number of mask edges, we define a small number of regions in the observed rest frame that are relatively free from the strongest telluric absorption and night sky emission features.  For each object, these unmasked regions were checked for any major contamination (such as stellar flux set to zero). If any such contamination was found, the entire unmasked region was discarded. On the green chip, the four unmasked regions are 1.5875--1.5968, 1.6050--1.6065, 1.6150--1.6215,  and 1.6240--1.6330~$\mu$m. On the red chip, the three unmasked regions are  1.6505--1.6552, 1.6556--1.6650, and  1.685--1.690~$\mu$m.

%A cross-correlation is performed for each combination of chip and template spectrum, and the peak of the resulting cross-correlation function (CCF) is found and fit with a Gaussian function plus  a linear background. The FWHM of that gaussian is determined by the convolution of the line widths in the object and template spectra. Each object-template pair results in four independent correlations (for the combinations of the two combined spectra for each star) per chip, which are averaged. Because both the object and template spectra should be in the stellar rest frame, the CCF should peak at relative velocity of 0~\kms. CCF peaks that are more than 5~\kms\ from zero are discarded from the average. We also discard measurements for which the cross-correlation peak height is $<$ 0.4. Low peak heights indicate poor matching between the star and template. 

To map FWHM to $v \sin i$,  we broaden each of the template spectra with a rotation kernel for a range of $v \sin i$ between 2--30~km s$^{-1}$ in increments of 2~km s$^{-1}$,  from 30--60~km s$^{-1}$ in increments of 5~km s$^{-1}$, and from 60--90~km s$^{-1}$ in increments of 10~km s$^{-1}$.  The rotational kernel is given by Equation 17.12 of \cite{gray_phot}, using a limb darkening parameter ($\epsilon$) of 0.25, which is appropriate for the NIR H band \citep{claret12}. These broadened templates are used as object spectra and run through the same pipeline as the program stars. For a given kernel  $v \sin i$ we averaged the resulting FWHM of all the broadened spectra for each combination of chip and un-broadened template.  This allowed us to create a mapping from FWHM to $v \sin i$ for each template and chip.  The true $v \sin i$ modeled by the artificially broadened template includes both the intrinsic rotation of the template stars and the applied rotation kernel. To account for the intrinsic rotation, we used 4.5~km s$^{-1}$ as an estimate of the intrinsic stellar rotation, and set the $v \sin i$ of the artificially broadened spectra to be the quadrature sum of the intrinsic $v \sin i$ and the kernel $v \sin i$. To apply this mapping to the object spectra, we fit a 6th-order polynomial to each relationship. The relationship for the blue chip was very different from that of the red and green chips, and we decided to exclude that chip in this paper. Although the mapping to $v \sin i$ varies little between each template, especially at low $v \sin i$, we still mapped each combination of chip and template independently. 

Before using the polynomials to map the FWHM of the object spectra to $v \sin i$, we first apply a quality cut requiring that the height of the cross-correlation peak be at least 0.4 and that the center of the cross-correlation peak be within $\pm$ 5 km s$^{-1}$ of zero, since all apStar spectra should be in the stellar rest frame. These cuts should remove correlations for which the largest peak is in fact a noise peak.  The FWHM measurements making this cut are mapped to $v \sin i$, and we compute an upper and lower $v \sin i$ by mapping to $v \sin i$ the average FWHM plus and minus the standard deviation in the FWHM, respectively.  These are used for upper and lower errors in $v \sin i$.  In all of these cases, small FWHM that would otherwise map to negative values are set to zero. Next, we average together all the $v \sin i$ measurements for a given chip. We propagate the high and low $v \sin i$ errors by taking the quadrature sum of each. We also compute the standard deviation in the $v \sin i$ values ($\sigma_{\rm chip}$) as a measure of systematic errors arising from the different templates. Finally, we take a weighted average of the two individual chip $v \sin i$'s to compute the final $v \sin i$, denoted ($v \sin i$)$_f$.  The weights are given by  $w=1/\sigma_{\rm chip}^2$. We set a floor of $\sigma_{\rm chip}$ to 0.5~km s$^{-1}$, to avoid giving overly high weights. The final error in the $v \sin i$ is given by the quadrature sum of the formal errors and the measured standard deviation.

The two methods, as discussed above, independently measure $v \sin i$ of the M dwarfs from Year 1, and are directly compared in Figure \ref{vsini_comp}. For most stars with $v~\sin~i$~$>$ 9 km s$^{-1}$, the measurements derived from two methods are within 2-$\sigma$ of the 1:1 line. However, for $v \sin i$ below 9 km s$^{-1}$ we find that method 2 either underestimates velocities or method 1 overestimates them for some of the stars. Nonetheless, most of the slow rotators are within 1-$\sigma$ of the 1:1 line. 

\begin{figure}[t!]
\includegraphics[scale=0.55]{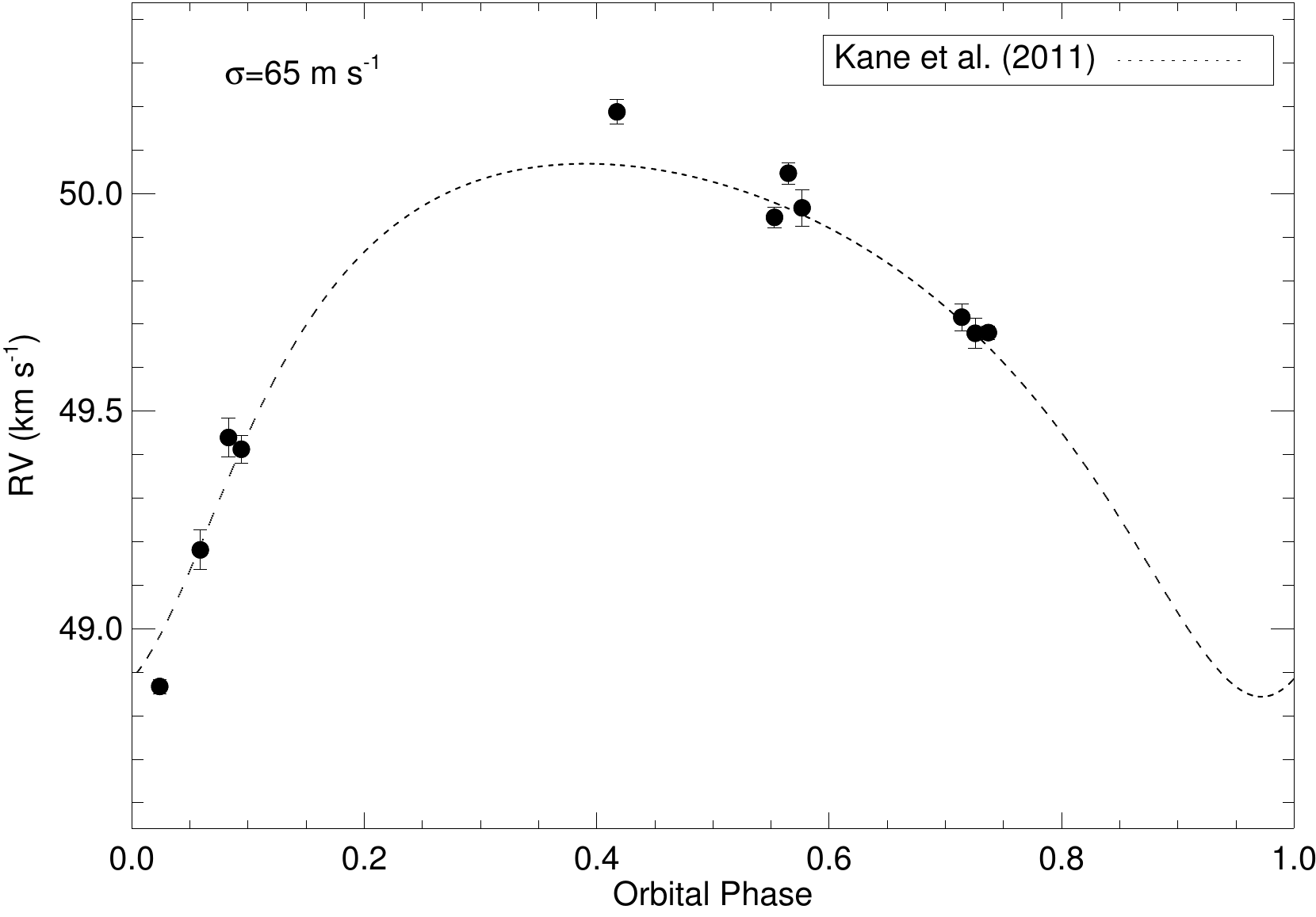}
\caption{APOGEE relative RV measurements compared to the published orbit for HD114762.\label{rvHD}}
\end{figure}

Both methods rely on a comprehensive library of templates. Method 1, uses the BT-Settl models \citep{allard97} at the finest grid these atmospheric models provide. However, Method 2 is limited by a small number of observed templates. This is a consequence of finding few $v \sin i$ templates that overlapped with pre-designed APOGEE fields. Preliminary APOGEE spectra pipeline analysis suggests that all of the observed templates have T$_{\rm{eff}}$ between 3300-3500 K while their V--J colors confirm them to be early M dwarfs. The same analysis done on our sample indicates a T$_{\rm{eff}}$ range of 3500 K - 2700 K with the mean T$_{\rm{eff}}$ of 3300 K. However, there are $\sim$ 70 stars with T$_{\rm{eff}}$ below 3100 K. As the chemical composition of M dwarfs changes faster than the effective temperature \citet{reiners12}, early M dwarf templates are likely to give erroneous measurements for late-M dwarfs. For completeness and due to lack of a comprehensive library of observed templates in our sample, we report $v \sin i$ measurements as derived by Method 1. These measurements are listed in Table 1 along with their errors. Figure \ref{vsini_plot} plots $v \sin i$ measurements derived from Method 1 (blue filled circles) as a function of V--J color along with those from literature. Upper limits on slow rotators of 4 km s$^{-1}$ are indicated by downward arrows. For comparison we also plot literature $v \sin i$ values (gray filled circles).

\begin{figure}[t!]
\centering
\includegraphics[scale=0.5]{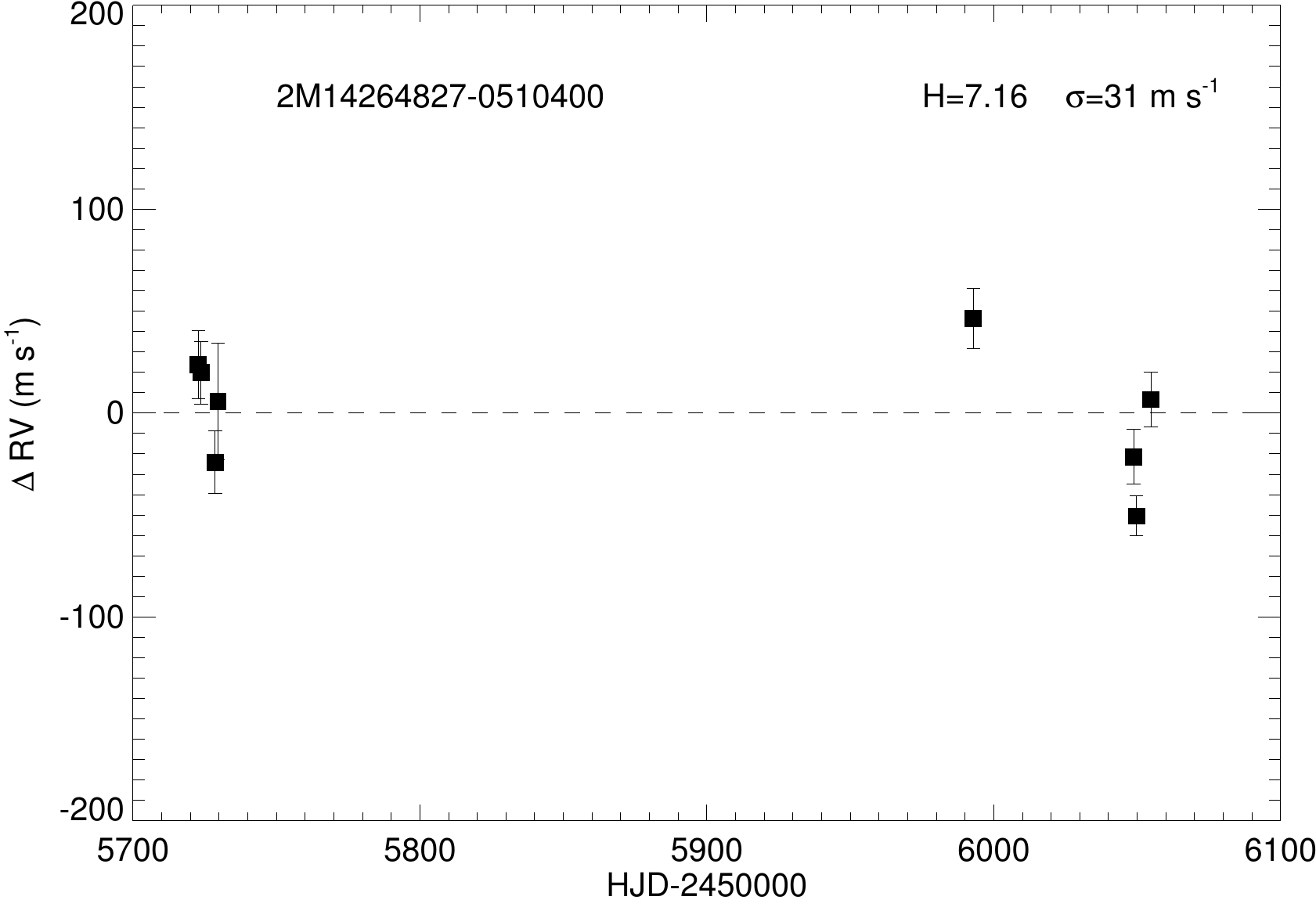}
\includegraphics[scale=0.5]{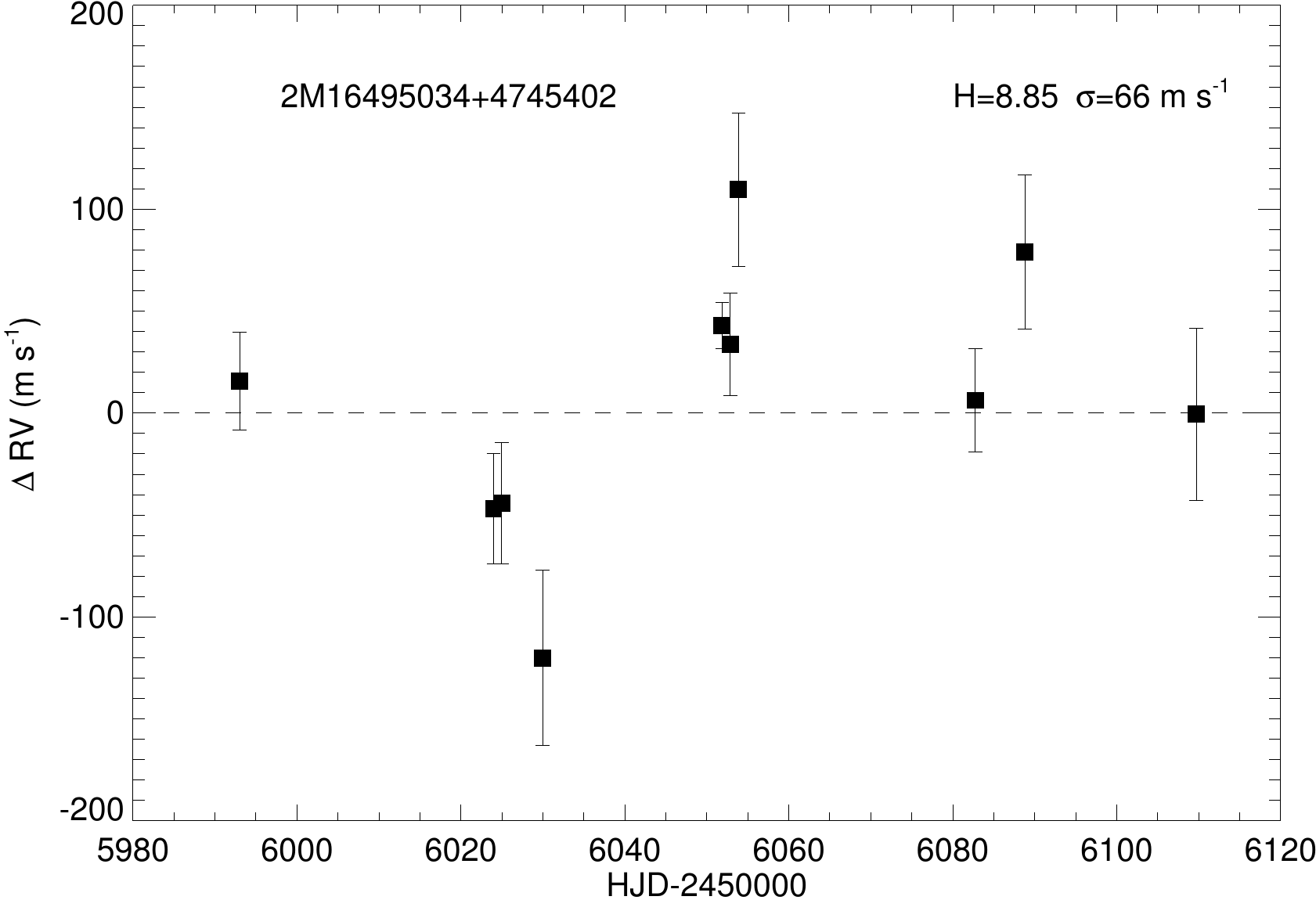}
\caption{APOGEE relative RVs for two low-mass stars. 2M1426+0510 (top panel) is a K0 star with a known planetary companion, resulting in a RV semi-amplitude of 7.32 m s$^{-1}$ at an orbital period of 3.41 years \citep{howard10}.\label{rv_panel}}
\end{figure}

\subsection{Relative Radial Velocities using Telluric Modelling}

APOGEE is, in principle, very well suited for making precise RV measurements of low-mass stars because of the LSF stability intrinsic to a fiber-fed instrument housed in a vacuum enclosure. While the modest resolution of the instrument precludes precision at the level achieved by instruments designed specifically for RV planet searches at optical wavelengths, such as HARPS \citep{mayor2003}, precision better than 50 m s$^{-1}$ should be possible for bright targets. Given the low masses of our targets, this is sufficient to detect a wide range of companions, including giant planets with short orbital periods.

Precise RV instruments generally rely on one of two strategies for calibrating the pixel-to-wavelength scale of the instrument and monitoring short- and long-term instrumental drifts: an emission line source coupled via an optical fiber to a stabilized instrument or an absorption reference gas cell in the telescope beam prior to the entrance of a slit spectrograph. The primary wavelength calibration of the APOGEE data relies on a combination of ThAr and UNe lamps \citep{redman11, redman12}, but the numerous telluric absorption features present in the APOGEE spectra can act as a wavelength reference, providing a secondary wavelength calibration that is obtained simultaneously with the stellar spectra. The bulk motion of the atmosphere causes instability in telluric absorption features at the 5--10 m s$^{-1}$ level (e.g. \citealt{figueira2012} and references therein). Given the resolution of APOGEE and the typical S/N of the spectra, we expect the RV precision limitations imposed by the intrinsic stability of the telluric lines to be a factor of a few smaller than the limitations imposed by the intrinsic information content of the spectra and the stability of the spectrograph. 

To estimate small corrections to wavelength solutions derived from the emission line lamps alone, we forward model the APOGEE spectra using a technique similar to the one used in \citet{blake2010} and \citet{rodler2012}, which are both based on the data analysis strategy outlined in \citet{butler1996}. We model spectra spanning 850 pixels from the green chip, covering the approximate wavelength range 1.598 to  1.616 $\mu$m. While this is less than a third of the total spectral coverage of APOGEE, this region is rich with telluric and stellar spectral features and also suffers less from the undersampling of the LSF found at bluer wavelengths. Following \citet{blake2010}, we forward model the spectra as the product of a stellar template and telluric model convolved with the spectrograph LSF. For a stellar template, we use the apStar spectra, the weighted combination of all the individual spectra of each object, produced by the APOGEE pipeline. We deconvolve this apStar template with the LSF estimated by the APOGEE pipeline (Nidever et al., 2013) using a Jansson technique based on that used in \citet{butler1996}. Given enough APOGEE epochs over a wide enough range of barycentric velocity, this technique can effectively average out errors due to imprecise telluric modeling and residuals from the subtraction of bright sky lines. We note that constructing a template in this way makes all of our resulting RV measurements fundamentally differential and also relies on the assumption that any intrinsic stellar RV variations are small compared to the barycentric motions and the APOGEE pixels (1 pixel~$\approx$~5~km~s$^{-1}$)

In the spectral region we are focusing on, there are prominent telluric absorption features due to CO$_2$. We calculate a high-resolution telluric transmission model appropriate for average conditions at Apache Point Observatory using the radiative transfer code described in \citet{blake2011}. Our model for the APOGEE spectra has four free parameters: one optical depth scale factor for the telluric model, two for a linear wavelength correction to the APOGEE pipeline wavelength solutions, and one for stellar velocity. We use the LSFs provided by the APOGEE pipeline. We fit the individual apCframe spectra, several of which are obtained at each epoch as part of a dither set. The model is first generated at a resolution seven times the APOGEE resolution, then interpolated to the APOGEE wavelength scale while explicitly integrating over the extent of each detector pixel. We use the AMOEBA downhill simplex method \citep{nelder65} to fit the model to each apCframe spectrum and find the best-fit values for the four free parameters by minimizing $\chi^2$. Prior to fitting each spectrum we mask out regions known to have strong OH emission lines. Finally, we decorrelate the RV estimates from the forward modeling process against the measured skewness of the APOGEE pipeline LSFs for each apCframe spectrum. We found that this correction improves the scatter of the resulting RV measurements by up to 30~m~s$^{-1}$.

In Figures \ref{rvHD} and \ref{rv_panel} we show examples of RV precision, including a star known to be RV stable to better than 10 m s$^{-1}$ and the giant planet companion to the star HD114762 \citep{latham1989, kane2011}. This was observed serendipitously by APOGEE as a telluric standard, and was independently recovered as a signal of a possible sub-stellar companion. It serves to demonstrate that NIR radial velocities even with a survey instrument are at the level of precision to enable discovery of giant planets. The telluric modeling techniques presented here are still being developed, improved, and more precise radial velocities deriving from this analysis will be presented in future work.

\subsection{Barycentric Radial Velocities}

Barycentric radial velocities for each star are derived in the final step of the pipeline reduction process. The detailed description of this RV determination is given in Nidever et al. (2013). Here, we summarize the steps. Radial velocity is determined by cross-correlating a normalized observed spectrum with a library of synthetic BT-Settl models that span the effective temperature and $\log{g}$ range of M dwarfs with metallicities between -4.0 and 0.3. First, a model template is selected through cross-correlating each model template from the library with the observed target. The best template is chosen through $\chi^2$ minimization. This template is then cross-correlated with the observed spectra of the target and velocity is estimated by fitting the CCF with a Gaussian plus a linear fit. Finally, a barycentric correction is applied to this velocity measurement. The standard deviation of the RV measurements from multiple observations is $\sim$130 m s$^{-1}$. The absolute RVs of stars from our sample are listed in Table 1. For the sake of consistency these velocities are the same as those released as part of the pipeline output with the DR10 release. The column of $\sigma$-RV in Table 1 shows the RMS scatter in the measured radial velocities for stars that have measurements at three or more epochs. 

Stars that show RV variation larger than 1 km s$^{-1}$ or result in CCFs with distinct peaks are identified as potential binaries. These candidate binaries are listed in Table 4 along with the number of observations and their infrared magnitudes. Orbital parameters, mass ratios and other analysis of these systems will be presented in separate papers, along with additional radial velocities and high-resolution spectra.

\subsection{Physical Parameters of M dwarfs}

Important improvements in stellar atmospheric models of low-mass stars have resulted from the availability of new atomic line profile data and the inclusion of dust and clouds in the models. Atomic line profile data becomes especially important in situations where line blanketing and broadening are crucial, and stellar models incorporating these updated atomic data give a much improved representation of the details of the line shapes in optical and NIR spectra of cool dwarfs \citep{rajpurohit12}. The recent suite of synthetic BT-Settl models include dust and clouds in their computation of stellar atmospheres and therefore provide a good fit to the observed stellar spectrum. These model atmospheres are computed with the PHOENIX code assuming hydrostatic equilibrium and convection using mixing length theory \citep{prandtl26} with a mixing length of $1/H_{\rm{p}}=2.0$ according to results of radiation hydrodynamics \citep{ludwig06}. The models are calculated using spherically symmetric radiative transfer, departure from LTE for all elements up to iron, the latest solar abundances by \citep{asplund09,caffau11}, equilibrium chemistry, a database of the latest opacities and thermochemical data for atomic and molecular transitions, and monochromatic dust condensates and refractory indexes. Grains are assumed spherical and non-porous, and their Rayleigh and Mie reflective and absorptive properties are considered. The diffusive properties of grains are treated based on 2-D radiation hydrodynamic simulations, including forsterite cloud formation to account for the feedback effects of cloud formation on the mixing properties of these atmospheres \citep{freytag10}. They are distributed via the PHOENIX web simulator\footnote{http://phoenix.ens-lyon.fr/Grids/BT-Settl/AGSS2009/}.

\begin{figure}
\includegraphics[scale=0.50]{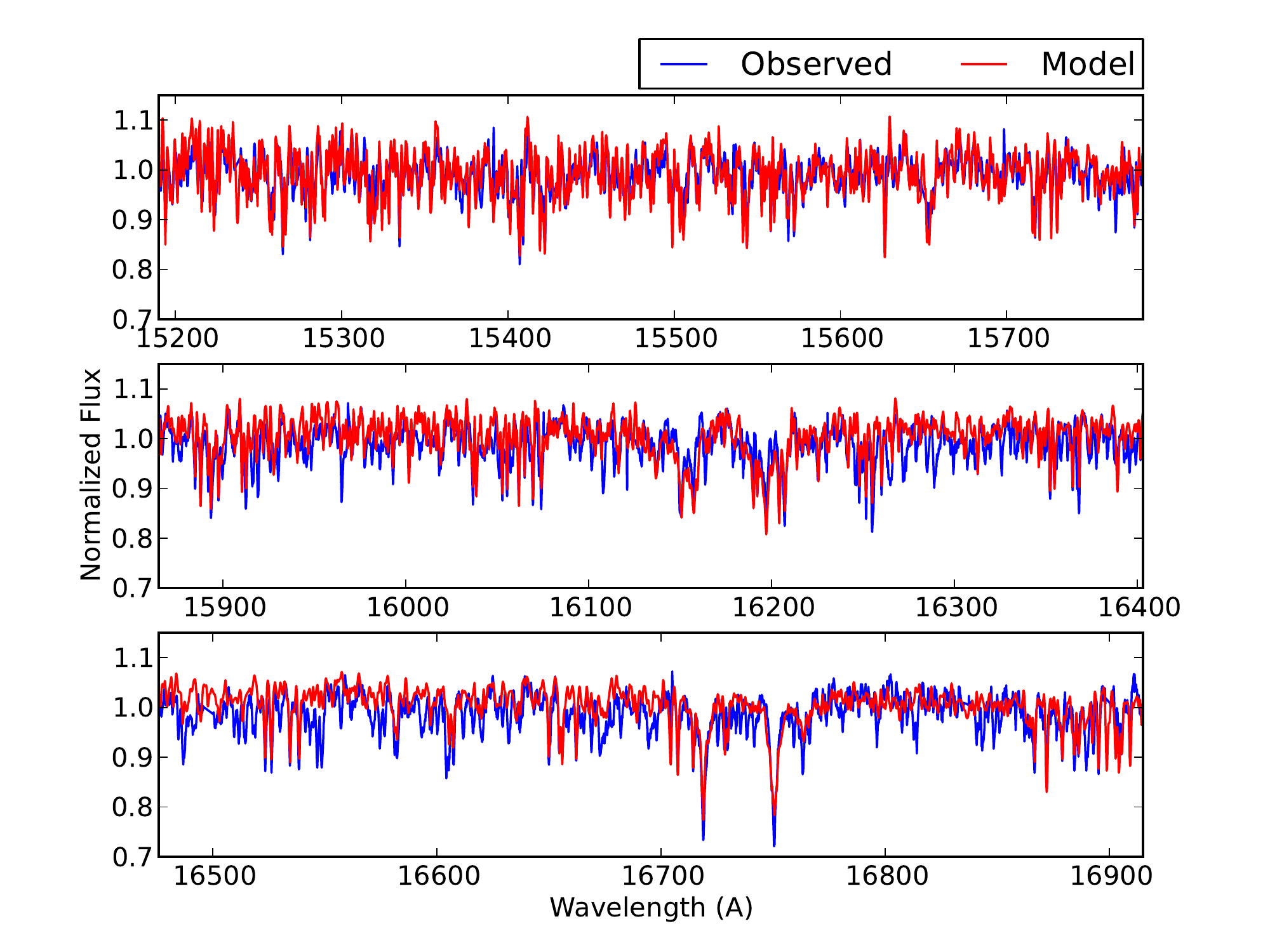} \\
\caption{Comparison between the observed spectrum (blue) and the model (red) for one of our M dwarf targets. The stellar parameters for this star are listed in Table \ref{table3}. The BT-Settl model that fits best has the following parameter: T$_{\rm{eff}}$ = 3000 K; $\log{g}$ = 5.0; Fe/H = 0.3. The three panels illustrate the wavelength coverage of three chips: the blue chip (top panel), the green chip (center panel) and the red chip (bottom panel).\label{atmos_model}}
\end{figure}

We selected four APOGEE M dwarfs that have also been observed by us with the IRTF \citep{terrien12a} and for which we have derived empirical H-band low-resolution metallicities. We used these stars and performed a detailed analysis between BT-Settl models and the spectra. We compute synthetic spectra over the entire spectral range of interest using a model grid described as follows: T$_{\rm{eff}}$ from 2000 K to 4000 K with 100 K steps, log g = 5.0 and 5.5, [M/H] = -2.0 dex to +0.5 dex with 0.5 dex steps. We convolve the synthetic spectra with a Gaussian kernel at the APOGEE resolution and then rebin to match the pixel sampling of the observations. As a first step, we calculate a $\chi^2$ goodness of fit statistic by comparing the observed spectra with the grids of synthetic spectra. This allows us to estimate effective temperature. Next, assuming this effective temperature we use the most prominent lines in the spectra, the atomic Al lines at 16718$\rm{\AA}$ and 16750$\rm{\AA}$, to constrain the other parameters (log g and [M/H]). The derived parameters are given in Table \ref{table3}. The comparison between the observed spectrum (blue) and the best fit model (red) is shown in Figure \ref{atmos_model} for 2M19081153+2839105.

We derive stellar parameters using our improved atmosphere models and spectroscopic information covering the given NIR range. Metallicity and gravity are determined from specific spectral features such as \ion{Al}{1}, \ion{Fe}{1} and \ion{Ca}{1}, whereas effective temperatures are constrained from the overall shape of the spectra via the following steps: (1) a first $\chi^2$ minimization is performed on the overall spectra considering effective temperature, metallicity, and gravity as free parameters. It gives a first guess for the parameter space of each component; (2) we look for those specific spectral features that are mainly sensitive to metallicity or gravity to refine these two parameters; (3) we fixed these parameters to perform another $\chi^2$ minimization and derive effective temperature. At each step we check that the resulting value is not sensitive to changes in the values of the other parameters

\begin{figure*}[t]
\includegraphics[scale=0.6]{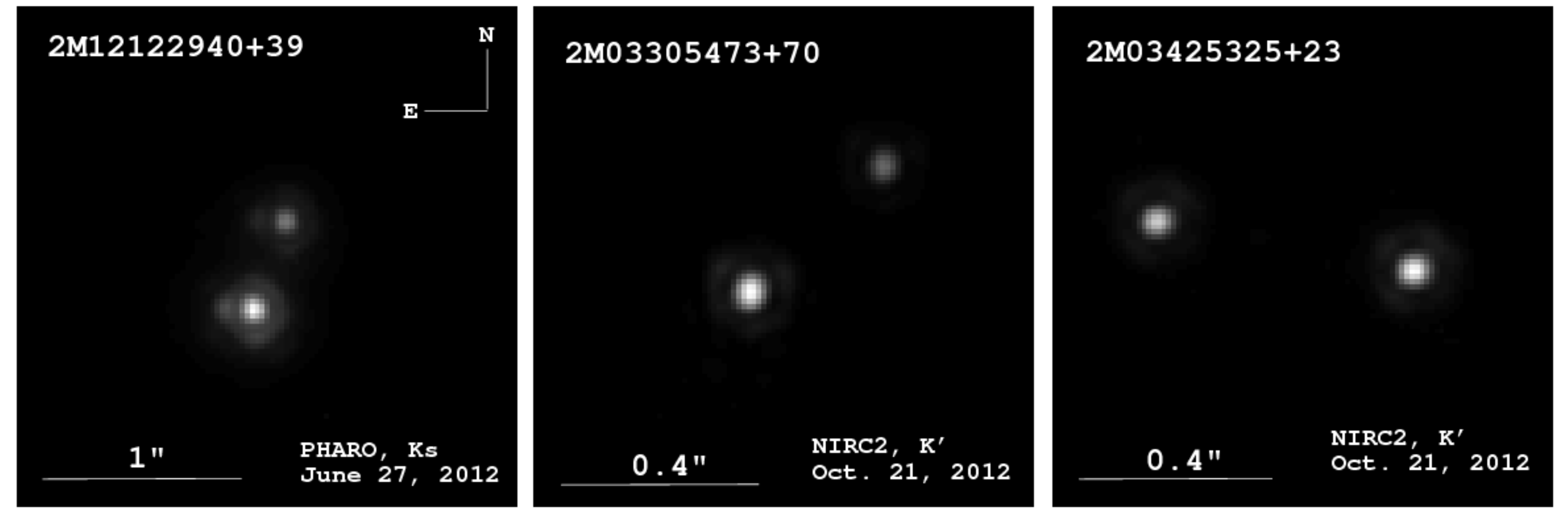}
\caption{ Adaptive optics images of three APOGEE M-dwarf targets taken at Palomar (left) and Keck (middle, right). Our diffraction-limited observations help assess the multiplicity of sources by probing separations complementary to RV observations.\label{ao_imaging}}
\end{figure*}

The synthetic spectra reproduce very well most of the spectral features like \ion{Fe}{1}, \ion{Ca}{1}, \ion{Al}{1}. Some discrepancies remains in the strengths of some atomic lines, which are either too strong or weak in the model, such as \ion{Fe}{1} (15267.02~\AA, 15237.7~\AA, 15335.00~\AA, 15964.87~\AA), and  \ion{Ca}{1} (16197.04~\AA). This is mainly due to the fact that the models have to use some incomplete or approximate input physics, such as broadening damping constants and uncertain oscillator strengths, for some lines and molecular bands. Also, the BT-Settl models use general atomic damping constants according to \cite{unsold68} with a correction factor to the van der Waals widths of 2.5 \citep{valenti96}, and van der Waals broadening of molecular lines with generic widths according to \citet{homeier03}. Furthermore, the BT-Settl models do not have a good handle on modeling the H$_{2}$O bands. This discrepancy is clearly seen in panel 3 of Figure \ref{atmos_model}

Recent progress has been made in the development of empirical calibrations that allow estimation of [Fe/H] for M dwarfs to a precision of 0.12 -- 0.15 dex \citep[][]{rojas-ayala12, rojas-ayala10,terrien12a}. \cite{terrien12a} developed such a calibration that uses the equivalent widths of specific regions in R $\sim$2000 $H$ band M dwarf spectra \citep{terrien12a}. This relation is calibrated using a set of 22 M dwarf companions to FGK stars having well-constrained metallicities and by assuming the individual systems are coeval. Column 3 in Table \ref{table3} lists the metallicity of three M dwarfs observed by APOGEE and that have had their metallicity determined using this technique using spectra from the IRTF. The three methods agree with each other within the errors. The more comprehensive comparison study of the metallicity determination through empirical spectroscopic relationships and stellar atmospheric modeling will be explored in a subsequent paper. 

The features used for the $H$ band calibration are K I (1.52 $\mu$m) and Ca I (1.62 $\mu$m). These features fall inside the spectral region covered by APOGEE, and so should be amenable to a similar empirical calibration strategy as employed with R $\sim$ 2000 spectra used in \citet{terrien12a}. The results of this empirical metallicity calibration for APOGEE observations of M dwarfs will be presented in a subsequent paper.

\subsection{AO Imaging of M dwarfs}
We have initiated an extensive campaign to collect adaptive optics (AO) imaging of a significant sample of M dwarfs in the APOGEE RV survey in an effort to also detect binaries at wider separations. A number of factors make this attractive and practical:
\begin{enumerate}
\item Because they are intrinsically faint, M dwarfs offer less demanding contrast requirements compared to solar-type stars for the direct imaging detection of low-mass companions at small angular separations.
\item M dwarfs in close proximity to the Sun ($d\lesssim20$ pc) are (nevertheless) sufficiently bright to serve as their own natural guide star.
\item AO instruments offer diffraction-limited performance at near-infrared wavelengths, facilitating the detection of faint (and red) companions whose black-body radiation peaks in the $\lambda \approx 1.0-5.0\mu$m range.
\item By concentrating companion light into a compact and locally-intense point-spread-function, the sensitivity of AO observations benefits from an increased signal-to-noise ratio compared to seeing-limited or speckle observations.
\end{enumerate}

Large aperture telescopes equipped with AO imagers, such as Palomar \citep{bouchez09}, Keck \citep{wiz00}, and the Large Binocular Telescope \citep{esposito12}, provide sensitivity to stellar companions of any mass using only seconds of integration time. High-contrast observations, such as those using AO in combination with a coronagraph and/or point-spread function subtraction \citep{marois06}, are sensitive to brown dwarfs over essentially all masses and ages with $\approx$ 1 hr integration times, even at sub-arcsecond separations \citep{crepp12a}. 

The scientific motivation for combining precision RV measurements with AO observations is equally compelling. By combining two complementary observing techniques, it is possible to place strong constraints on the presence of both short-period and long-period companions around each star. A joint Doppler and imaging survey will enable detailed studies of stellar multiplicity at the low-mass end of the main sequence. Further, the frequency of brown dwarf companions to low-mass stars is still unknown in a large fraction of the parameter space explored by our survey. For instance, \citet{metchev09} have quantified the occurrence rate of brown dwarfs orbiting FGK stars in the 29-1590 AU range, but similar studies for M-dwarf primaries have only recently commenced \citep{bowler12}.

In addition to discovering a plethora of short-period companions, APOGEE's multiplexing capabilities will also reveal systems that exhibit long-term Doppler accelerations indicating the presence of unseen wide-separation companions. RV "trends" act as a signpost to identify promising AO imaging follow-up targets \citep{crepp12b}. In the case of a direct detection, multi-epoch AO imaging and continued Doppler measurements can ultimately lead to the construction of three-dimensional orbits and calculation of dynamical masses \citep{crepp12b}. Such mass "benchmark" systems may in turn be used to explicitly calibrate theoretical atmospheric models and theoretical evolutionary models of cool dwarfs. Finally, in the case of a non-detection, it is possible to place strong constraints on the mass and period of putative companions \citep{rodigas11,montet13}.  

Motivated by these factors, we have commenced AO observations of M-dwarf targets in the APOGEE target list starting with the brightest sources. Several candidate companions have been identified using the PALM-3000 AO system at Palomar \citep{bouchez09} and PHARO camera \citep{hayward01} as shown in Figure \ref{ao_imaging}. Given their proximity to the Sun, M dwarfs in our sample generally have a high proper motion. With ~10 mas astrometry precision, consecutive observations separated by only several months may be used to unambiguously determine whether each candidate shares a common proper motion with its host star. First epoch and follow-up AO measurements are on-going. Figure \ref{ao_imaging} gives examples of some of the M dwarf binary candidates that have been discovered through AO imaging. Table \ref{table4} lists their angular separations, position angles, and delta magnitudes.  

\section{Discussion $\&$ Future Prospects}
The APOGEE M dwarf survey described here will produce a catalog of multi-epoch RV measurements of more than 1400 low-mass stars. These RV measurements are derived from high-S/N, high-resolution (R$\sim22500$) H-band spectra gathered as part of an ancillary science program of the main SDSS-III APOGEE survey. These observations will be used to identify individual low-mass spectroscopic binaries, and possibly even short period giant planets. The APOGEE spectra routinely provide velocity precision better than 100 m s$^{-1}$, and we have shown that for bright targets, a detailed analysis of the spectra and the telluric absorption lines they contain can result in precision better than 50 m s$^{-1}$. At the same time, these spectra can also be modeled to measure projected rotational velocities, $v\sin{i}$, and chemical composition. We present $v\sin{i}$ measurements for over 200 M stars, a significant increase in the total number of available rotation measurements for low-mass stars. Another outcome of this survey is the determination of metallicity through empirical measurements and with the use of stellar atmosphere models such as the BT-Settl models. Combining the results of different analyses of the APOGEE M dwarf spectra will provide a wealth of information about the structure and evolution of the lowest mass stars. 

One of the primary goals of this survey is to quantify the rate of occurrence of companions to M dwarfs over a wide range of separations and mass ratios. This complete picture of M dwarf multiplicity will place important observational constraints on theoretical models of the formation of stars at the bottom of the main sequence. We will carry out a joint analysis of wide-separation companions identified through AO imaging and small-separation companions detected as spectroscopic binaries in the APOGEE data. With over 1400 targets in our RV sample, we expect to detect a large number of binaries, significantly more than any previous single survey. Kinematic measurements derived from these RV data will also be important for placing the local population of M dwarfs in the context of Galactic stellar populations. 

\begin{figure}[t!]
\includegraphics[scale=0.55]{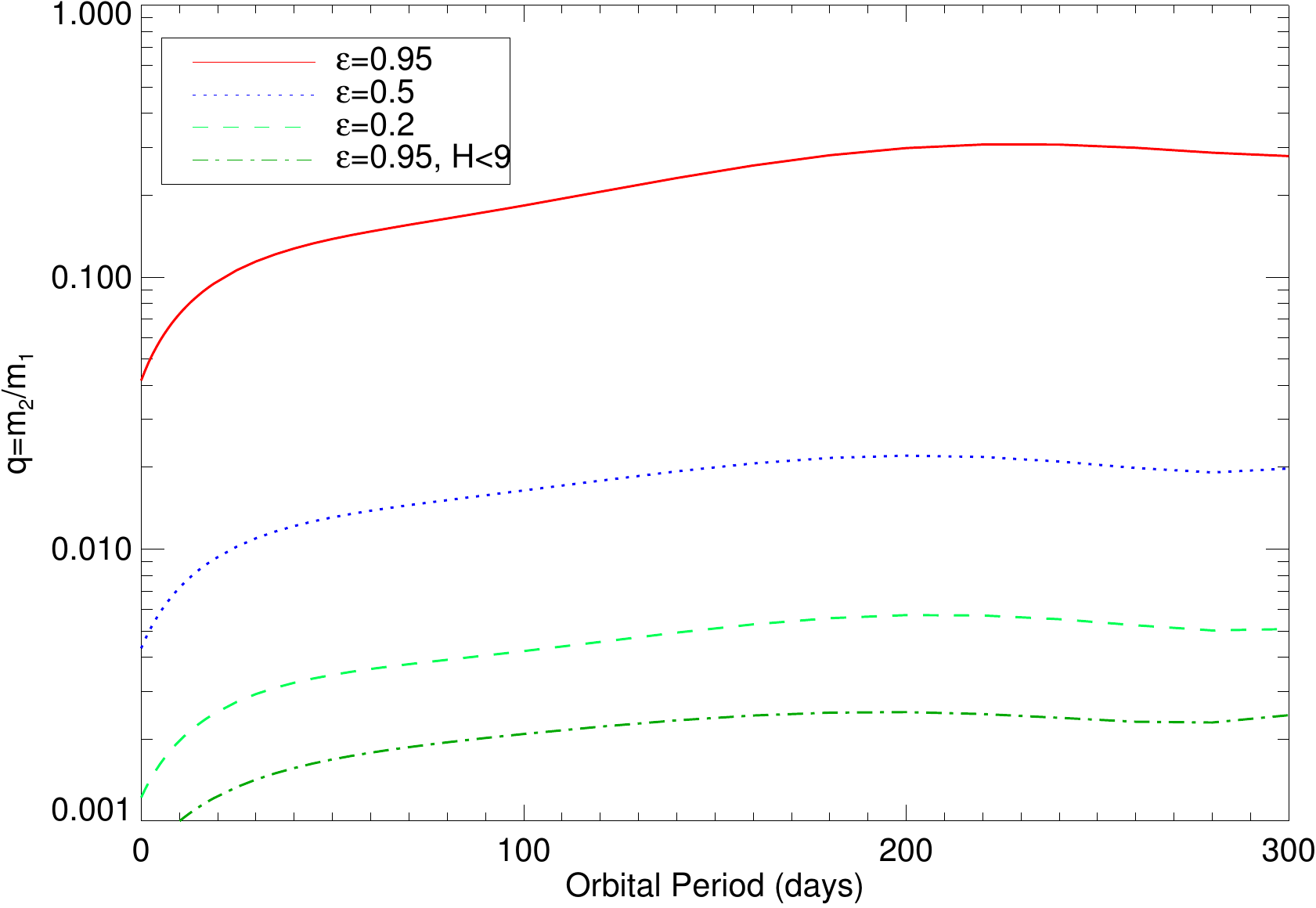} \\
\caption{Results of a simulation of detection efficiency. Areas above each line correspond to detection efficiency above a given $\epsilon$. For example, we expect $95\%$ sensitivity to companions with mass ratios $q>0.25$ and orbital periods less than 300 days (red, solid line).\label{detefficiency}}
\end{figure}

We have carried out a Monte Carlo simulation to estimate our expected sensitivity to short-period companions with a range of masses. We use the actual observational cadence for over 200 M dwarfs observed by APOGEE so far and a simple model for RV precision as a function of magnitude ($\sigma_{RV}=[40, 40, 66, 110, 180, 300]~\rm{m \; s^{-1}}$ for $H=[7,8,9,10,11,12]$~mag) to simulate $10^5$ APOGEE RV surveys. We assume Gaussian noise for stars drawn randomly from the actual $H$ magnitude distribution (Figure 3) and inject RV variations resulting from Keplerian orbits. We assume each star has a mass of $0.4~M_{\sun}$ and consider mass ratios $q=m_{2}/m_{1}$ in the range 0.001 to 1.0 and orbital periods from 1 to 300 days. We select random inclinations, orbital phases, and orbital orientations, and set eccentricity to $e=0$ for periods less than 10 days and $e$ uniformly distributed between 0 and 0.8 for longer periods. For each simulated set of RV measurements of a star we calculate $\chi^{2}$ assuming the null hypothesis of RVs consistent with no variation and count the injected Keplerian orbit as "detected" if the probability of $\chi^{2}$ greater than or equal to the given value is less than 1$\%$. We average the percentage of detections, $\epsilon$, in bins of $q$ and orbital period and show the results of this simulation in Figure \ref{detefficiency}. Contours corresponding to $95\%$, $50\%$ and $20\%$ detection efficiency are shown. 

Our simulations indicate that the survey will have nearly complete sensitivity to companions with masses greater than $0.08~M_{\sun}$ and orbital periods less than 300 days. At orbital periods less than 10 days, we have excellent sensitivity to companions down to $25~M_{\rm{J}}$. For stars brighter than $H$=9 (N=235), we are sensitive to $0.5~M_{\rm{J}}$ companions out to orbital periods of 40 days. With this sensitivity and the published occurrence rates for M dwarf binaries and giant planets orbiting M dwarfs, we expect to detect dozens of new spectroscopic binary systems and place strong limits on the frequency of giant planet companions around M dwarfs of all spectral types. 

Given its large sample size and excellent RV precision, the APOGEE M Dwarf Radial Velocity Survey will provide a rich data set useful for addressing a number of important outstanding questions about low-mass stars. This survey will generate a large, homogeneous sample of M dwarf measurements, and when completed in 2014, will represent an unprecedented resource useful for studying stellar populations, binary statistics, and measuring physical stellar parameters. At the same time, the physical stellar parameters derived from the APOGEE observations will greatly benefit future planet search surveys such as HPF \citep{mahadevan12} and CARMENES \citep{quirrenbach10} by providing detailed information about chemical composition of low-mass stars that are found to host planets. This first paper in a series serves to describe the target selection and present rotational velocities and radial velocities derived from over 1000 individual spectra of more than 200 M dwarfs. The data described in this paper become publicly available as part of the SDSS-III DR10 data release in July, 2013.

%Final paragraph.

\acknowledgments

This work was partially supported by funding from the Center for Exoplanets and Habitable Worlds. The Center for Exoplanets and Habitable Worlds is supported by the Pennsylvania State University, the Eberly College of Science, and the Pennsylvania Space Grant Consortium. We acknowledge support from NSF grant AST 1006676 and AST 1126413 in our pursuit of precision radial velocities in the NIR. This research has made use of the SIMBAD database, operated at CDS, Strasbourg, France. This publication makes use of data products from the Two Micron All Sky Survey, which is a joint project of the University of Massachusetts and the Infrared Processing and Analysis Center/California Institute of Technology, funded by the National Aeronautics and Space Administration and the National Science Foundation. The authors are visiting Astronomer at the Infrared Telescope Facility, which is operated by the University of Hawaii under Cooperative Agreement no. NNX-08AE38A with the National Aeronautics and Space Administration, Science Mission Directorate, Planetary Astronomy Program. JRC acknowledges support from NASA Origins grant NNX13AB03G. This research has made use of NASA's Astrophysics Data System. 

This work was based on observations with the SDSS 2.5-meter telescope. Funding for SDSS-III has been provided by the Alfred P. Sloan Foundation, the Participating Institutions, the National Science Foundation, and the U.S. Department of Energy Oﬃce of Science. The SDSS-III web site is http://www.sdss3.org/. SDSS-III is managed by the Astrophysical Research Consortium for the Participating Institutions of the SDSS-III Collaboration including the University of Arizona, the Brazilian Participation Group, Brookhaven National Laboratory, University of Cambridge, Carnegie Mellon University, University of Florida, the French Participation Group, the German
Participation Group, Harvard University, the Instituto de Astroﬁsica de Canarias, the Michigan State/Notre Dame/JINA Participation Group, Johns Hopkins University, Lawrence Berkeley National Laboratory, Max Planck Institute for Astrophysics, Max Planck Institute for Extraterrestrial Physics, New Mexico State University, New York University, Ohio State University, Pennsylvania State University, University of Portsmouth, Princeton University, the Spanish Participation Group, University of Tokyo, University of Utah, Vanderbilt University, University of Virginia, University of Washington, and Yale University.

{\it Facilities:} \facility{SDSS, IRTF}.

%********************************** BIBLIOGRAPHY ***************************************************************

%********************************** TABLES ***************************************************************
\newpage
\begin{center}
\tabletypesize{}
\scriptsize
\begin{longtable}{|c|r|r|r|c|c|r|r|r|r|c|}
\caption{Targets observed during the first year of the APOGEE M Dwarf Survey. The table includes infrared magnitudes, absolute RVs, projected rotational velocities and estimated errors on absolute RV and $v \sin i$. } \\
\hline
\textbf{} & \textbf{} & \textbf{} & \textbf{} & \textbf{} & \textbf{} & \textbf{} & \textbf{} & \textbf{} & \textbf{} & \textbf{}\\
\textbf{Object} & \textbf{J mag} & \textbf{H mag} & \textbf{K mag} & \textbf{Visits} & \textbf{Visits} & \textbf{RV} & \textbf{$\sigma$-RV} & \textbf{$v \sin i$} & \textbf{$\sigma$-$v \sin i$} & \textbf{Binary}\\
\textbf{(2MASS ID)} & \textbf{} & \textbf{} & \textbf{} & \textbf{Planned} & \textbf{Yr 1.} & \textbf{(km s$^{-1}$)} & \textbf{(km s$^{-1}$)} & \textbf{(km s$^{-1}$)} & \textbf{(km s$^{-1}$)} & \textbf{}\\
\textbf{} & \textbf{} & \textbf{} & \textbf{} & \textbf{} & \textbf{} & \textbf{} & \textbf{} & \textbf{} & \textbf{} & \textbf{}\\
\hline
\endfirsthead
\multicolumn{4}{c}%
{\tablename\ \thetable\ -- \textit{Continued from previous page}} \\
\hline
\textbf{} & \textbf{} & \textbf{} & \textbf{} & \textbf{} & \textbf{} & \textbf{} & \textbf{} & \textbf{} & \textbf{} & \textbf{}\\
\textbf{Object} & \textbf{J mag} & \textbf{H mag} & \textbf{K mag} & \textbf{Visits} & \textbf{Visits} & \textbf{RV} & \textbf{$\sigma$-RV} & \textbf{$v \sin i$} & \textbf{$\sigma$-$v \sin i$} & \textbf{Binary}\\
\textbf{(2MASS ID)} & \textbf{} & \textbf{} & \textbf{} & \textbf{Planned} & \textbf{Yr. 1} & \textbf{(km s$^{-1}$)} & \textbf{(km s$^{-1}$)} & \textbf{(km s$^{-1}$)} & \textbf{(km s$^{-1}$)} & \textbf{}\\
\textbf{} & \textbf{} & \textbf{} & \textbf{} & \textbf{} & \textbf{} & \textbf{} & \textbf{} & \textbf{} & \textbf{} & \textbf{}\\
\hline
\endhead

\hline \multicolumn{4}{l}{\textit{Continued on next page}} \\
\endfoot
\hline
\endlastfoot

\hline \multicolumn{3}{|r|}{{Continued on next page}} \\ \hline
\endfoot

\hline \hline
\endlastfoot
2M00034394+8606422	&	12.307	&	11.738	&	11.485	&	5	&	3	&	10.80		&	0.28	&	13.20	&	1.50   	&  \nodata  	\\
2M00085424+6716518	&	11.674	&	11.131	&	10.836	&	5	&	12	&	10.07		&	0.19	&	27.60	&	1.40   	&  \nodata   \\
2M00110613+7202521	&	12.247	&	11.660	&	11.414	&	4	&	12	&	$-$7.64		&	0.24	&	4.20	&	1.70   	&  \nodata   \\
2M00131578+6919372	&	8.556	&	7.984	&	7.746	&	4	&	12	&	13.57		&	0.17	&	6.00	&	0.20   	&  \nodata   \\
2M00184520+7040399	&	12.185	&	11.620	&	11.389	&	4	&	12	&	0.05		&	0.15	&	$<4.00$	& \nodata   &  \nodata   \\
2M00185999+5836527	&	11.839	&	11.264	&	11.009	&	6	&	12	&	$-$7.75		&	0.38	&	$<4.00$	& \nodata   &  \nodata   \\
2M00222083+8619567	&	11.564	&	11.018	&	10.675	&	5	&	3	&	$-$36.70	&	0.10	&	8.10	&	0.60   	&  \nodata   \\
2M00234573+5353478	&	12.060	&	11.483	&	11.179	&	3	&	12	&	5.24		&	0.47	&	10.10	&	1.20   	&  \nodata   \\
2M00251480+6225017	&	12.306	&	11.777	&	11.461	&	6	&	12	&	29.55		&	0.26	&	6.90	&	1.60   	&  \nodata   \\
2M00251602+5422547	&	11.775	&	11.215	&	10.819	&	3	&	12	&	$-$63.27	&	0.20	&	9.20	&	0.40   	&  \nodata   \\
2M00252974+5412240	&	11.656	&	11.018	&	10.790	&	4	&	12	&	$-$10.34	&	0.36	&	$<4.00$	& \nodata	&  \nodata     \\
2M00255540+5749320	&	11.607	&	11.035	&	10.762	&	6	&	12	&	$-$26.05	&	0.17	&	5.40	&	0.40   	&  \nodata   \\
2M00274401+5330504	&	11.486	&	10.891	&	10.606	&	4	&	12	&	$-$19.48	&	0.11	&	8.90	&	2.10   	&  \nodata   \\
2M00321574+5429027	&	9.387	&	8.827	&	8.570	&	4	&	12	&	$-$14.75	&	0.13	&	5.70	&	0.40   	&  \nodata   \\
2M00331747+6327504	&	12.284	&	11.758	&	11.489	&	5	&	12	&	$-$16.55	&	0.10	&	6.90	&	0.50   	&  \nodata   \\
2M00333033+7054422	&	11.277	&	10.691	&	10.452	&	4	&	12	&	$-$78.93	&	0.37	&	6.70	&	0.80   	&  \nodata   \\
2M00350487+5953079	&	11.039	&	10.401	&	10.166	&	6	&	12	&	0.40		&	0.17	&	7.40	&	1.10   	&  \nodata   \\
2M00391896+5508132	&	10.021	&	9.519	&	9.236	&	4	&	12	&	$-$29.76	&	0.18	&	5.70	&	0.70   	&  \nodata   \\
2M01094190+8612302	&	12.026	&	11.327	&	11.089	&	5	&	3	&	$-$52.30	&	0.08	&	9.90	&	0.70   	&  \nodata   \\
2M01195227+8409327	&	9.855	&	9.314	&	9.025	&	5	&	3	&	$-$41.01	&	0.09	&	5.80	&	0.90   	&  \nodata   \\
2M01333457+6220188	&	11.500	&	10.916	&	10.645	&	2	&	3	&	$-$44.19	&  \nodata	&	4.30	&	2.80   	&  \nodata   \\
2M01382032+6246328	&	12.496	&	11.966	&	11.669	&	2	&	3	&	$-$47.80	&  \nodata	&	$<4.00$	& \nodata 	&  \nodata     \\
2M01422966+8601414	&	10.865	&	10.241	&	10.022	&	5	&	3	&	$-$14.24	&	0.19	&	11.00	&	0.70   	&  \nodata   \\
2M02040123+4855372	&	10.935	&	10.368	&	10.076	&	4	&	3	&	31.28		&	0.15	&	$<4.00$	& \nodata   &  \nodata   \\
2M02081366+4949023	&	12.043	&	11.476	&	11.136	&	4	&	3	&	$-$63.77	&	0.27	&	10.10	&	0.80   	&  \nodata   \\
2M02085359+4926565	&	8.423	&	7.811	&	7.584	&	4	&	3	&	\nodata	    & \nodata	&	\nodata	& \nodata   & SB1  \\
2M02144781+5334438	&	11.814	&	11.272	&	11.036	&	1	&	3	&	39.39		&  \nodata	&	5.80	&  \nodata 	&   \nodata    \\
2M02170514+5434535	&	10.482	&	10.043	&	9.781	&	1	&	3	&	$-$43.43	&  \nodata	&	$<4.00$	& \nodata 	&   \nodata    \\
2M02210709+5457023	&	11.812	&	11.158	&	10.907	&	1	&	3	&	27.04		&  \nodata	&	5.60	&  \nodata 	&   \nodata    \\
2M02250239+6006593	&	11.817	&	11.244	&	10.999	&	3	&	3	&	$-$15.23	&	0.32	&	$<4.00$	& \nodata 	&   \nodata    \\
2M02595985+6514530	&	11.901	&	11.274	&	11.010	&	1	&	3	&	16.67		&  \nodata	&	8.70	&  \nodata 	&   \nodata    \\
2M03042484+6641064	&	12.381	&	11.801	&	11.511	&	1	&	3	&	$-$8.29		&  \nodata	&	7.20	&  \nodata 	&   \nodata    \\
2M03104313+3854093	&	11.607	&	10.975	&	10.748	&	4	&	3	&	$-$19.35	&	0.38	&	5.80	&	0.30   	&   \nodata  \\
2M03140624+5728568	&	11.528	&	10.873	&	10.615	&	2	&	3	&	27.81		&  \nodata	&	7.90	&	1.50   	&   \nodata  \\
2M03150691+4016231	&	12.204	&	11.635	&	11.355	&	4	&	3	&	54.22		&	0.18	&	6.10	&	1.20   	&   \nodata  \\
2M03152943+5751330	&	11.121	&	10.533	&	10.271	&	2	&	3	&	76.71		&  \nodata	&	6.70	&	0.20   	&   \nodata  \\
2M03220945+7047486	&	10.913	&	10.284	&	10.007	&	5	&	3	&	$-$17.52	&	0.10	&	5.90	&	1.90   	&   \nodata  \\
2M03224269+7932066	&	11.040	&	10.491	&	10.220	&	11	&	24	&	$-$0.96		&	0.21	&	9.80	&	0.80   	&   \nodata  \\
2M03264491+7108292	&	10.877	&	10.268	&	10.022	&	5	&	3	&	46.42		&	0.12	&	5.90	&	0.30   	&   \nodata  \\
2M03300766+4711159	&	11.610	&	11.018	&	10.737	&	8	&	12	&	13.66		&	0.46	&	5.90	&	1.50   	&   \nodata  \\
2M03305285+5627325	&	10.191	&	9.531	&	9.287	&	2	&	3	&	15.67		&  \nodata	&	4.20	&	2.10   	&   \nodata  \\
2M03305473+7041145	&	9.487	&	8.938	&	8.675	&	5	&	3	&	21.15		&	0.07	&	4.70	&	1.70   	&   \nodata  \\
2M03334490+7054444	&	11.337	&	10.742	&	10.460	&	5	&	3	&	5.56		&	0.25	&	$<4.00$	& \nodata   &   \nodata  \\
2M03355099+7140275	&	10.340	&	9.725	&	9.446	&	5	&	3	&	5.35		&	0.20	&	8.10	&	0.50   	&  \nodata   \\
2M03425325+2326495	&	10.202	&	9.545	&	9.316	&	3	&	3	&	35.55		&	0.04	&	12.70	&	0.50   	&  \nodata   \\
2M03425617+2515528	&	11.513	&	11.061	&	10.706	&	3	&	3	&	106.26		&	0.11	&	8.80	&	1.50   	&  \nodata   \\
2M03431519+5006558	&	11.486	&	10.878	&	10.628	&	5	&	12	&	28.98		&	0.31	&	6.70	&	0.90   	&  \nodata   \\
2M03441913+7126195	&	11.905	&	11.271	&	11.049	&	5	&	3	&	9.68		&	0.41	&	$<4.00$	& \nodata   &  \nodata   \\
2M03443389+7125059	&	11.916	&	11.283	&	11.036	&	5	&	3	&	11.32		&	0.38	&	6.80	&	1.50   	&  \nodata  \\
2M03494426+8027289	&	11.834	&	11.272	&	10.994	&	11	&	24	&	\nodata	    & \nodata	&	\nodata	& \nodata   & SB1  \\
2M03504584+8026223	&	11.226	&	10.669	&	10.411	&	11	&	24	&	5.40		&	0.31	&	$<4.00$	& \nodata   &  \nodata    \\
2M03515527+2356547	&	12.314	&	11.787	&	11.587	&	3	&	3	&	$-$47.91	&	0.19	&	$<4.00$	& \nodata   &  \nodata    \\
2M04030024+5238026	&	12.016	&	11.504	&	11.239	&	6	&	12	&	4.83		&	0.14	&	5.90	&	0.70   	&  \nodata    \\
2M04063732+7916012	&	10.034	&	9.486	&	9.194	&	11	&	24	&	1.36		&	0.14	&	5.20	&	2.00   	&  \nodata    \\
2M04081814+5127550	&	11.724	&	11.148	&	10.862	&	6	&	12	&	29.10		&	0.21	&	$<4.00$	& \nodata   &  \nodata    \\
2M04104186+5124238	&	11.817	&	11.223	&	10.951	&	6	&	12	&	22.76		&	0.19	&	$<4.00$	& \nodata	&  \nodata      \\
2M04125880+5236421	&	8.773	&	8.248	&	7.915	&	6	&	12	&	\nodata	    & \nodata	&	\nodata	& \nodata   &  SB1 \\
2M04234272+5605429	&	12.279	&	11.691	&	11.400	&	6	&	12	&	21.05		&	0.13	&	7.70	&	2.50   	&  \nodata \\
2M04281703+5521194	&	12.514	&	11.912	&	11.648	&	6	&	12	& 	\nodata	    & \nodata	&	\nodata	& \nodata  	&  SB2   \\
2M04293734+5455319	&	12.083	&	11.445	&	11.167	&	6	&	12	&	23.81		&	0.11	&	7.30	&	2.10   	&  \nodata    \\
2M04314006+5501136	&	11.948	&	11.304	&	10.973	&	6	&	12	&	22.92		&	0.11	&	10.80	&	1.50   	&  \nodata    \\
2M04422854+5818015	&	10.754	&	10.186	&	9.946	&	6	&	12	&	27.26		&	0.13	&	7.20	&	1.00   	&  \nodata    \\
2M04480390+5718092	&	12.525	&	11.939	&	11.623	&	6	&	12	&	$-$32.00	&	0.25	&	6.80	&	0.80   	&  \nodata    \\
2M04480517+3732402	&	12.504	&	11.888	&	11.609	&	6	&	12	&	41.08		&	0.44	&	5.60	&	1.60   	&  \nodata    \\
2M04542406+3830436	&	11.346	&	10.797	&	10.529	&	6	&	12	&	13.95		&	0.21	&	9.00	&	1.20   	&  \nodata    \\
2M04572425+2149303	&	11.958	&	11.348	&	11.050	&	2	&	3	&	35.99		&  \nodata	&	$<4.00$	& \nodata   &  \nodata    \\
2M05011792+4204125	&	10.676	&	10.047	&	9.808	&	3	&	3	&	20.06		&	0.13	&	7.60	&	0.80   	&  \nodata    \\
2M05011802+2237015	&	10.161	&	9.591	&	9.232	&	2	&	3	&	31.56		&  \nodata	&	8.80	&	0.30   	&  \nodata    \\
2M05014944+2127375	&	11.591	&	10.962	&	10.718	&	2	&	3	&	$-$39.94	&  \nodata	&	5.10	&	3.20   	&  \nodata    \\
2M05030563+2122362	&	9.750	&	9.165	&	8.888	&	2	&	3	&	49.93		&  \nodata	&	16.80	&  \nodata 	&  \nodata      \\
2M05062444+2230199	&	12.037	&	11.470	&	11.158	&	2	&	3	&	$-$31.73	&  \nodata	&	7.50	&	3.80   	&  \nodata    \\
2M05104796+2410125	&	10.187	&	9.588	&	9.333	&	6	&	12	&	31.07		&	0.31	&	$<4.00$	& \nodata 	&  \nodata      \\
2M05172841+2531214	&	11.750	&	11.130	&	10.888	&	5	&	12	&	$-$26.76	&	0.11	&	7.80	&	2.40   	&  \nodata    \\
2M05210188+3425119	&	11.879	&	11.319	&	11.022	&	3	&	3	&	6.54		&	0.18	&	7.80	&	3.40   	&  \nodata    \\
2M05255532+3528438	&	12.468	&	11.900	&	11.626	&	3	&	3	&	81.45		&	0.04	&	10.30	&	0.40   	&  \nodata    \\
2M05320969+2754534	&	12.562	&	11.976	&	11.613	&	4	&	12	&	$-$22.81	&	0.18	&	13.20	&	1.90   	&  \nodata    \\
2M05424060+6134403	&	11.424	&	10.835	&	10.574	&	3	&	3	&	$-$46.26	&	0.15	&	6.50	&	0.50   	&  \nodata    \\
2M05445320+3018018	&	11.658	&	11.014	&	10.733	&	5	&	12	&	32.55		&	0.20	&	7.30	&	0.90   	&  \nodata \\
2M05470907$-$0512106 &	10.039	&	9.514	&	9.177	&	3	&	3	&	$-$54.88	&	0.20	&	6.40	&	1.70   	&  \nodata \\
2M06034934+3019208	&	9.735	&	9.128	&	8.845	&	6	&	12	&	14.48		&	0.17	&	6.90	&	1.00   	&  \nodata \\
2M06070493+1403109	&	11.421	&	11.063	&	10.768	&	6	&	12	&	50.68		&	0.16	&	6.80	&	1.20   	&  \nodata \\
2M06115599+3325505	&	10.163	&	9.591	&	9.345	&	6	&	12	&	\nodata		&	\nodata	&	\nodata	& \nodata  	&  SB2  \\
2M06205178+3145134	&	11.362	&	10.775	&	10.462	&	6	&	12	&	83.61		&	0.15	&	9.60	&	0.70   	&  \nodata \\
2M06213640+3222390	&	11.041	&	10.433	&	10.155	&	5	&	12	&	\nodata		&	\nodata	&	\nodata	& \nodata   &  SB1 \\
2M06215046+1554140	&	11.389	&	10.817	&	10.553	&	1	&	3	&	35.65		&  \nodata	& \nodata	&  \nodata 	&  \nodata      \\
2M06293433+1742419	&	11.755	&	11.195	&	10.905	&	6	&	12	&	20.88		&	0.22	&	6.30	&	0.80   	&  \nodata    \\
2M06293962+1723109	&	12.481	&	11.976	&	11.678	&	6	&	12	&	38.98		&	0.33	&	$<4.00$	& \nodata   &  \nodata    \\
2M06320207+3431132	&	10.692	&	10.143	&	9.864	&	1	&	3	&	10.57		&  \nodata	&	$<4.00$	& \nodata 	&  \nodata      \\
2M06362535+1830520	&	11.014	&	10.485	&	10.193	&	6	&	12	&	$-$36.60	&	0.17	&	6.40	&	0.20   	&  \nodata    \\
2M06481555+0326243	&	11.555	&	10.943	&	10.585	&	5	&	12	&	17.41		&	0.11	&	16.60	&	1.10   	&  \nodata    \\
2M07015902+0456273	&	11.923	&	11.411	&	11.141	&	5	&	12	&	128.60		&	0.27	&	7.60	&	1.30   	&   \nodata   \\
2M07104573+0619385	&	12.335	&	11.869	&	11.569	&	13	&	12	&	33.17		&	0.61	&	11.10	&	2.30   	&   \nodata   \\
2M07125105+3707340	&	11.114	&	10.596	&	10.265	&	3	&	3	&	24.32		&	0.49	&	20.40	&	1.30   	&   \nodata   \\
2M07140394+3702459	&	11.976	&	11.252	&	10.838	&	3	&	3	&	40.03		&	0.11	&	12.80	&	0.50   	&   \nodata   \\
2M07382463+0948552	&	12.288	&	11.687	&	11.457	&	3	&	3	&	$-$7.53		&	0.24	&	5.10	&	1.60   	&   \nodata   \\
2M07404603+3758253	&	11.787	&	11.275	&	11.002	&	7	&	3	&	\nodata	    & \nodata	&	\nodata	&	\nodata &   SB1 \\
2M07454991+3716280	&	11.589	&	10.997	&	10.712	&	7	&	3	&	$-$22.57	&	0.14	&	6.70	&	1.00   	&  \nodata    \\
2M08160527+3028386	&	12.508	&	11.910	&	11.610	&	6	&	12	&	$-$35.94	&	0.29	&	7.90	&	1.20   	&  \nodata    \\
2M08501918+1056436	&	11.282	&	10.675	&	10.407	&	3	&	3	&	32.97		&	0.09	&	9.20	&	2.10   	&  \nodata    \\
2M08505983+1130493	&	12.374	&	11.795	&	11.533	&	3	&	3	&	51.93		&	0.27	&	11.00	&	1.50   	&  \nodata    \\
2M08584396+3727389	&	12.137	&	11.585	&	11.341	&	2	&	3	&	4.19		&  \nodata	&	6.20	&	0.80   	&  \nodata    \\
2M09031291+3739077	&	12.578	&	11.956	&	11.709	&	2	&	3	&	16.24		&  \nodata	&	9.80	&	2.00   	&  \nodata    \\
2M09120266+3636314	&	12.230	&	11.647	&	11.342	&	2	&	3	&	28.72		&  \nodata	&	$<4.00$	& \nodata  	&  \nodata    \\
2M09301445+2630250	&	8.866	&	8.284	&	8.020	&	2	&	3	&	20.84		&  \nodata	&	6.70	&	1.50   	&  \nodata    \\
2M09321296+2745465	&	11.722	&	11.101	&	10.799	&	2	&	3	&	20.14		&  \nodata	&	10.40	&	0.80   	&  \nodata    \\
2M09391173+3837561	&	11.921	&	11.335	&	11.037	&	2	&	3	&	20.73		&  \nodata	&	9.00	&	1.10   	&  \nodata    \\
2M10150280+0140004	&	11.595	&	10.986	&	10.767	&	1	&	3	&	42.04		&  \nodata	&	$<4.00$	& \nodata 	&  \nodata      \\
2M10162955+0318375	&	10.859	&	10.264	&	10.007	&	1	&	3	&	$-$2.49		&  \nodata	&	$<4.00$	& \nodata 	&  \nodata      \\
2M10164852+0041408	&	12.357	&	11.807	&	11.481	&	1	&	3	&	41.03		&  \nodata	&	9.50	&  \nodata 	&  \nodata      \\
2M10245696+1721531	&	11.966	&	11.402	&	11.064	&	3	&	3	&	41.00		&	0.05	&	9.80	&	1.80   	&  \nodata    \\
2M10304397+1624207	&	11.783	&	11.174	&	10.875	&	3	&	3	&	38.00		&	0.01	&	5.10	&	2.20   	&  \nodata    \\
2M10345281+3911043	&	10.863	&	10.228	&	9.989	&	2	&	3	&	33.81		&  \nodata	&	$<4.00$	& \nodata 	&  \nodata      \\
2M10393463+3809488	&	11.616	&	11.051	&	10.794	&	2	&	3	&	13.86		&  \nodata	&	6.10	&	0.60   	&  \nodata    \\
2M10464238+1626144	&	11.497	&	10.835	&	10.617	&	3	&	3	&	\nodata 	& \nodata	& \nodata	& \nodata  	&  SB2  \\
2M10540048+1606059	&	11.606	&	11.077	&	10.731	&	3	&	3	&	23.93		&	0.05	&	10.10	&	0.80   	&  \nodata    \\
2M11005043+1204108	&	10.676	&	10.117	&	9.782	&	3	&	3	&	$-$10.07	&	0.13	&	26.50	&	0.80   	&  \nodata    \\
2M11014478+1227162	&	10.986	&	10.425	&	10.162	&	3	&	3	&	27.14		&	0.13	&	11.60	&	0.80   	&  \nodata    \\
2M11045698+1026411	&	9.368	&	8.712	&	8.492	&	3	&	3	&	8.45		&	0.15	&	4.70	&	2.10   	&  \nodata    \\
2M11054316+1014093	&	8.643	&	8.049	&	7.797	&	3	&	3	&	$-$56.59	&	0.16	&	5.50	&	2.20   	&  \nodata    \\
2M11091225$-$0436249	&	8.201	&	7.595	&	7.330	&	1	&	3	&	10.23	&  \nodata	&	6.30	&  \nodata 	&  \nodata      \\
2M11285879+5205580	&	11.225	&	10.612	&	10.332	&	2	&	3	&	7.41		&  \nodata	&	2.90	&	0.50   	&  \nodata    \\
2M11463262+0130524	&	12.240	&	11.673	&	11.382	&	4	&	12	&	28.35		&	0.17	&	3.50	&	2.80   	&  \nodata    \\
2M11474074+0015201	&	8.991	&	8.399	&	8.098	&	4	&	12	&	7.02		&	0.32	&	5.60	&	1.40   	&  \nodata    \\
2M11564308+1639541	&	11.164	&	10.610	&	10.332	&	2	&	3	&	30.36		&  \nodata	&	$<4.00$	& \nodata 	&  \nodata      \\
2M11593840+1545481	&	11.423	&	10.784	&	10.504	&	2	&	3	&	18.61		&  \nodata	&	$<4.00$	& \nodata  	&  \nodata    \\
2M12003311+1814555	&	12.010	&	11.398	&	11.139	&	9	&	6	&	$-$4.40		&	0.42	&	7.20	&	2.90   	&  \nodata    \\
2M12033893+1527533	&	11.222	&	10.631	&	10.372	&	2	&	3	&	$-$11.23	&  \nodata	&	5.90	&	0.10   	&  \nodata    \\
2M12045611+1728119	&	9.793	&	9.183	&	8.967	&	2	&	3	&	\nodata	    & \nodata	&	\nodata	& \nodata  	&  SB2  \\
2M12045619+1757409	&	11.568	&	11.014	&	10.710	&	9	&	6	&	$-$2.36		&	0.22	&	4.30	&	3.10   	&  \nodata    \\
2M12054224+1844354	&	12.209	&	11.576	&	11.386	&	9	&	6	&	20.96		&	0.18	&	$<4.00$	& \nodata   &  \nodata    \\
2M12081374+1747306	&	12.433	&	11.925	&	11.654	&	9	&	6	&	15.11		&	0.25	&	$<4.00$	& \nodata   &  \nodata    \\
2M12110614+1846153	&	11.117	&	10.545	&	10.281	&	9	&	6	&	20.61		&	0.26	&	$<4.00$	& \nodata   &  \nodata    \\
2M12151947+0537224	&	12.324	&	11.767	&	11.454	&	1	&	3	&	29.78		&  \nodata	&	5.20	&  \nodata 	&    \nodata    \\
2M12200119+2633419	&	12.294	&	11.736	&	11.471	&	5	&	6	&	$-$8.66		&	0.49	&	$<4.00$	& \nodata   &    \nodata  \\
2M12211284+0031320	&	12.063	&	11.509	&	11.281	&	9	&	12	&	$-$21.77	&	0.21	&	$<4.00$	& \nodata  	&    \nodata  \\
2M12214665+0007205	&	11.842	&	11.255	&	10.969	&	9	&	12	&	$-$6.06		&	0.13	&	8.40	&	1.40   	&    \nodata  \\
2M12222073+0011020	&	11.788	&	11.273	&	11.018	&	9	&	12	&	31.09		&	0.20	&	9.60	&	0.50   	&    \nodata  \\
2M12232063+2529441	&	10.827	&	10.234	&	9.985	&	5	&	6	&	$-$12.01	&	0.08	&	$<4.00$	& \nodata   &    \nodata  \\
2M12243585+0001106	&	12.011	&	11.477	&	11.220	&	9	&	12	&	28.38		&	0.22	&	$<4.00$	& \nodata  	&   \nodata   \\
2M12244171+1405391	&	11.560	&	10.947	&	10.724	&	3	&	3	&	2.55		&	0.31	&	$<4.00$	& \nodata   &   \nodata   \\
2M12253159+0011354	&	11.649	&	11.105	&	10.805	&	9	&	12	&	20.96		&	0.36	&	$<4.00$	& \nodata   &   \nodata   \\
2M12260407+2730369	&	12.303	&	11.791	&	11.552	&	6	&	6	&	2.89		&	0.29	&	8.60	&	1.20   	&   \nodata   \\
2M12265737+2700536	&	10.197	&	9.607	&	9.320	&	6	&	6	&	$-$2.13		&	0.52	&	13.50	&	0.60   	&   \nodata   \\
2M13075012+1717471	&	12.327	&	11.742	&	11.486	&	16	&	12	&	11.27		&	0.25	&	$<4.00$	& \nodata   &   \nodata   \\
2M13085059+1622039	&	9.264	&	8.655	&	8.413	&	16	&	12	&	17.48		&	0.21	&	$<4.00$	& \nodata  	&    \nodata  \\
2M13332256+3620352	&	9.650	&	8.986	&	8.778	&	1	&	3	&	$-$23.02	&  \nodata	&	19.60	&  \nodata 	&    \nodata    \\
2M13345147+3746195	&	9.713	&	9.146	&	8.887	&	1	&	3	&	$-$3.77		&  \nodata	&	10.90	&  \nodata 	&    \nodata    \\
2M13410096+2711449	&	12.482	&	11.850	&	11.598	&	18	&	24	&	$-$19.25	&	0.24	&	$<4.00$	& \nodata   &    \nodata  \\
2M13451104+2852012	&	9.882	&	9.311	&	9.054	&	18	&	24	&	$-$11.41	&	0.18	&	$<4.00$	& \nodata   &    \nodata  \\
2M13455527+2723131	&	10.652	&	10.076	&	9.787	&	18	&	24	&	$-$43.14	&	0.18	&	6.40	&	0.80   	&    \nodata  \\
2M13491436+2637457	&	11.325	&	10.768	&	10.548	&	18	&	24	&	$-$9.47		&	0.13	&	$<4.00$	& \nodata 	&  \nodata      \\
2M13514938+4157445	&	9.894	&	9.272	&	9.024	&	2	&	3	&	$-$4.06		&  \nodata	&	8.10	&	0.60   	&  \nodata    \\
2M13532938+4416109	&	11.712	&	11.136	&	10.872	&	2	&	3	&	$-$6.42		&  \nodata	&	3.80	&	1.60   	&  \nodata    \\
2M13564148+4342587	&	11.709	&	11.043	&	10.650	&	2	&	3	&	$-$16.79	&  \nodata	&	18.80	&	0.40   	&  \nodata    \\
2M14035430+3008026	&	11.299	&	10.655	&	10.388	&	6	&	6	&	$-$39.48	&	0.26	&	7.60	&	1.00   	&  \nodata    \\
2M14045651+2831023	&	12.299	&	11.716	&	11.445	&	6	&	6	&	\nodata	    & \nodata	&	\nodata	& \nodata  	&  SB1  \\
2M14264827$-$0510400 &	7.470	&	7.160	&	7.060	&	5	&	6	&	$-$32.44	&	0.14	&	3.20	&	1.00   	&  \nodata    \\
2M14545704+3714558	&	10.423	&	9.769	&	9.531	&	1	&	3	&	$-$27.76	&  \nodata	&	5.10	&  \nodata 	&  \nodata      \\
2M14570070+3556471	&	9.758	&	9.089	&	8.890	&	1	&	3	&	$-$34.76	&  \nodata	&	3.00	&  \nodata 	&  \nodata      \\
2M14592508+3618321	&	10.257	&	9.647	&	9.377	&	1	&	3	&	$-$19.31	&  \nodata	&	9.70	&  \nodata 	&  \nodata      \\
2M15165576$-$0037116 &	9.960	&	9.382	&	9.105	&	1	&	3	&	$-$12.40	&  \nodata	&	11.60	&  \nodata 	&  \nodata      \\
2M15183842$-$0008235 &	9.342	&	8.645	&	8.499	&	1	&	3	&	$-$9.94		&  \nodata	&	54.50	&  \nodata 	&  \nodata      \\
2M15192613+0153284	&	12.133	&	11.610	&	11.329	&	10	&	12	& 	\nodata	    & \nodata	&	\nodata	& \nodata  	&  SB2  \\
2M16193140+5206469	&	11.438	&	10.913	&	10.687	&	1	&	3	&	$-$58.22	&  \nodata	&	5.70	&  \nodata 	&  \nodata      \\
2M16370146+3535456	&	11.135	&	10.545	&	10.240	&	7	&	12	&	$-$2.18		&	0.15	&	7.00	&	1.80   	&  \nodata    \\
2M16383835+3700273	&	11.367	&	10.798	&	10.498	&	7	&	12	&	14.59		&	0.17	&	8.30	&	1.50   	&  \nodata    \\
2M16440030+3721597	&	11.918	&	11.334	&	11.066	&	7	&	12	&	$-$0.66		&	0.12	&	6.80	&	0.40   	&  \nodata    \\
2M16451798+3645405	&	12.450	&	11.839	&	11.596	&	7	&	12	&	$-$76.21	&	0.31	&	8.30	&	1.20   	&  \nodata    \\
2M16454410+3605496	&	10.569	&	10.039	&	9.801	&	7	&	12	&	$-$9.12		&	0.05	&	5.30	&	1.20   	&  \nodata    \\
2M16495034+4745402	&	9.457	&	8.840	&	8.625	&	12	&	12	&	$-$35.63	&	0.15	&	6.70	&	0.90   	&  \nodata    \\
2M16500047+4820372	&	12.509	&	11.921	&	11.624	&	12	&	12	&	$-$12.34	&	0.20	&	13.00	&	1.20   	&  \nodata    \\
2M17204248+4205070	&	9.895	&	9.286	&	9.000	&	13	&	12	&	\nodata	    & \nodata	& \nodata	& \nodata   &  SB2 \\
2M17592886+0318233	&	9.474	&	8.830	&	8.633	&	3	&	3	&	12.26		&	0.13	&	11.00	&	4.60   	&  \nodata    \\
2M18055545+0316213	&	11.398	&	10.771	&	10.565	&	3	&	3	&	$-$14.06	&	0.31	&	6.30	&	1.50   	&  \nodata    \\
2M18215416$-$0700179 &	9.626	&	8.909	&	8.726	&	2	&	3	&	16.91		&  \nodata	&	$<4.00$	& \nodata   &  \nodata    \\
2M18244689$-$0620311 &	9.659	&	9.052	&	8.795	&	2	&	3	&	$-$20.96	&  \nodata	&	9.20	&	1.20   	&  \nodata    \\
2M18415473+0651285	&	9.921	&	9.336	&	9.050	&	2	&	3	&	$-$72.71	&  \nodata	&	6.50	&  \nodata 	&  \nodata      \\
2M18430881+0612150	&	12.175	&	11.813	&	11.688	&	2	&	3	&	20.47		&  \nodata	&	5.40	&	0.20   	&  \nodata    \\
2M18451027+0620158	&	7.656	&	7.043	&	6.806	&	2	&	3	&	$-$24.48	&  \nodata	&	7.10	&	0.20   	&  \nodata    \\
2M18452147+0711584	&	12.244	&	11.581	&	11.149	&	2	&	3	&	$-$46.19	&  \nodata	&	16.10	&	0.10   	&  \nodata    \\
2M18523373+4538317	&	10.493	&	9.937	&	9.673	&	3	&	2	&	$-$31.01	&	0.15	&	5.80	&	0.30   	&  \nodata    \\
2M18551497+4408590	&	12.172	&	11.596	&	11.302	&	3	&	2	&	10.90		&	0.27	&	4.90	&	2.70   	&  \nodata    \\
2M18562628+4622532	&	9.598	&	9.010	&	8.717	&	3	&	2	&	2.51		&	0.04	&	5.30	&	3.90   	&  \nodata    \\
2M19010098+4522386	&	11.539	&	10.902	&	10.639	&	3	&	2	&	$-$27.86	&	0.08	&	5.30	&	0.80   	&  \nodata    \\
2M19025183+4455465	&	10.058	&	10.017	&	10.016	&	3	&	2	&	\nodata	    & \nodata	& \nodata	& \nodata   &  SB1  \\
2M19034513+4536301	&	8.213	&	8.056	&	8.067	&	3	&	2	&	$-$0.85		&	0.05	&	4.20	&	0.60   	&  \nodata    \\
2M19051739+4507161	&	9.850	&	9.300	&	9.027	&	3	&	2	&	$-$48.22	&	0.23	&	2.40	&  \nodata 	&  \nodata      \\
2M19071270+4416070	&	10.447	&	9.855	&	9.559	&	2	&	2	&	$-$21.27	&  \nodata	&	7.10	&  \nodata 	&  \nodata      \\
2M19081153+2839105	&	10.581	&	9.973	&	9.719	&	7	&	12	&	30.57		&	0.14	&	22.70	&	0.90   	&  \nodata    \\
2M19081576+2635054	&	10.361	&	9.762	&	9.471	&	7	&	12	&	$-$3.84		&	0.12	&	2.80	&	0.20   	&  \nodata    \\
2M19084251+2733453	&	9.750	&	9.233	&	8.951	&	7	&	12	&	$-$5.35		&	0.09	&	5.70	&	0.20   	&  \nodata    \\
2M19121128+4316106	&	11.156	&	10.631	&	10.430	&	2	&	2	&	$-$110.63	&  \nodata	&	8.30	&	3.30   	&  \nodata    \\
2M19125504+4239370	&	10.315	&	9.731	&	9.543	&	2	&	2	&	$-$32.70	&  \nodata	&	$<4.00$	& \nodata   &  \nodata    \\
2M19142151+4309008	&	12.384	&	11.880	&	11.545	&	2	&	2	&	$-$77.36	&  \nodata	&	9.50	&  \nodata 	&  \nodata      \\
2M19173151+2833147	&	10.758	&	10.194	&	9.923	&	7	&	12	&	15.33		&	0.12	&	13.20	&	0.50   	&  \nodata    \\
2M19185898+3812236	&	11.876	&	11.222	&	11.024	&	3	&	3	&	$-$25.56	&	0.10	&	8.20	&	2.10   	&  \nodata    \\
2M19235494+3834587	&	12.534	&	12.087	&	11.973	&	3	&	3	&	\nodata	    & 	\nodata & \nodata	& \nodata  	&  SB2   \\
2M19241533+3638089	&	11.556	&	10.927	&	10.721	&	3	&	3	&	11.40		&	0.16	&	11.40	&	0.50   	&  \nodata    \\
2M19242742+2508341	&	12.386	&	11.789	&	11.545	&	3	&	12	&	40.36		&	0.49	&	$<4.00$	& \nodata   &  \nodata    \\
2M19265550+3844381	&	12.360	&	11.710	&	11.521	&	3	&	3	&	$-$15.88	&	0.31	&	12.10	&	2.50   	&  \nodata    \\
2M19271763+3913024	&	11.742	&	11.060	&	10.941	&	3	&	3	&	$-$23.24	&	0.18	&	19.60	&	0.10   	&  \nodata    \\
2M19272650+4758468	&	11.913	&	11.328	&	11.023	&	1	&	2	&	$-$23.33	&  \nodata	&	6.50	&  \nodata 	&  \nodata      \\
2M19294258+3733223	&	11.282	&	10.736	&	10.444	&	3	&	3	&	22.57		&	0.15	&	10.20	&	0.20   	&  \nodata    \\
2M19302029+3723437	&	11.939	&	11.310	&	11.109	&	3	&	3	&	23.06		&	0.35	&	13.50	&	0.40   	&  \nodata    \\
2M19321796+4747027	&	11.519	&	10.936	&	10.633	&	1	&	2	&	$-$6.61		&  \nodata	&	10.20	&  \nodata 	&  \nodata      \\
2M19322564+3917266	&	11.570	&	11.008	&	10.737	&	3	&	6	&	$-$28.64	&	0.02	&	5.60	&	0.70   	&  \nodata    \\
2M19330692+4005066	&	11.563	&	10.944	&	10.741	&	3	&	6	&	$-$20.10	&	0.07	&	9.00	&	0.10   	&  \nodata    \\
2M19332454+4515045	&	10.600	&	9.992	&	9.747	&	1	&	2	&	$-$5.82		&  \nodata	& \nodata	&  \nodata 	&  \nodata      \\
2M19333940+3931372	&	8.120	&	7.560	&	7.339	&	3	&	6	&		\nodata	& \nodata	&	\nodata	& \nodata   	&  SB2 \\
2M19335290+3900545	&	11.948	&	11.393	&	11.131	&	3	&	6	&	$-$24.52	&	0.04	&	$<4.00$	& \nodata   &  \nodata    \\
2M19335405+3938497	&	12.334	&	11.760	&	11.513	&	3	&	6	&	$-$46.05	&	0.10	&	5.70	&	1.10   	&  \nodata    \\
2M19395886+3950530	&	9.798	&	9.186	&	8.951	&	3	&	6	&	$-$18.45	&	0.47	&	$<4.00$	& \nodata   &  \nodata    \\
2M19412016+3912129	&	10.675	&	10.097	&	9.848	&	3	&	6	&	$-$49.72	&	0.06	&	4.60	&	1.70   	&  \nodata    \\
2M19420033+4038302	&	11.655	&	11.011	&	10.813	&	3	&	6	&	$-$45.15	&	0.23	&	11.00	&	0.30   	&  \nodata    \\
2M19421037+4005402	&	12.313	&	11.791	&	11.543	&	3	&	6	&	$-$76.57	&	0.06	&	7.40	&	1.70   	&  \nodata    \\
2M19430726+4518089	&	11.331	&	10.752	&	10.380	&	1	&	2	&	$-$2.05		&  \nodata	&	7.30	&  \nodata 	&  \nodata      \\
2M19443810+4720294	&	11.814	&	11.281	&	11.002	&	3	&	2	&	$-$2.53		&	0.26	& 	$<4.00$	& \nodata   &  \nodata    \\
2M19445931+4812415	&	12.557	&	11.937	&	11.735	&	3	&	2	&	$-$13.92	&	0.27	& 	$<4.00$	& \nodata   &  \nodata    \\
2M19450035+3911539	&	12.515	&	11.966	&	11.711	&	3	&	6	&	$-$10.33	&	0.45	&	4.40	&	2.40   	&  \nodata    \\
2M19450736+3947341	&	12.595	&	12.070	&	11.977	&	3	&	6	&	$-$52.11	&	0.34	&  	$<4.00$	& \nodata   &  \nodata    \\
2M19485718+5015245	&	11.184	&	10.606	&	10.389	&	2	&	2	&	3.24		&  \nodata	&	4.50	&	2.20   	&  \nodata    \\
2M19504779+4816290	&	11.390	&	10.880	&	10.662	&	3	&	2	&	$-$75.92	&	0.10	&	$<4.00$	& \nodata   	&  \nodata    \\
2M19510930+4628598	&	8.586	&	8.045	&	7.773	&	3	&	2	&	$-$11.38	&	0.19	&	19.00	&	0.40   	&  \nodata    \\
2M19541829+1738289	&	10.516	&	9.993	&	9.719	&	1	&	3	&	$-$40.84	&  \nodata	&	8.60	&  \nodata 	&  \nodata      \\
2M19544358+1801581	&	12.210	&	11.517	&	11.113	&	1	&	3	&	$-$16.08	&  \nodata	&	17.00	&  \nodata 	&  \nodata      \\
2M19560585+2205242	&	10.069	&	9.443	&	9.228	&	1	&	12	&	$-$8.62		&  \nodata	&	12.80	&  \nodata 	&  \nodata      \\
2M19591845+2040515	&	12.407	&	11.873	&	11.640	&	1	&	12	&	$-$33.51	&  \nodata	&	5.80	&  \nodata 	&  \nodata      \\
2M20184736+1901539	&	11.501	&	10.913	&	10.712	&	4	&	12	&	$-$35.46	&	0.19	&	5.10	&	0.90   	&  \nodata    \\
2M20264615+5250183	&	12.550	&	11.966	&	11.720	&	2	&	12	&	$-$18.72	&  \nodata	&	6.00	&  \nodata 	&  \nodata      \\
2M20351108+5426045	&	11.956	&	11.389	&	11.170	&	2	&	12	&	2.87		&  \nodata	& 	$<4.00$	& \nodata 	&  \nodata      \\
2M21003448+5156016	&	11.196	&	10.659	&	10.405	&	1	&	12	&	$-$3.51		&  \nodata	& 	$<4.00$	& \nodata 	&  \nodata      \\
2M21095447+4729043	&	11.487	&	10.900	&	10.633	&	4	&	24	&	$-$37.49	&	0.21	&	10.10	&	0.40   	&  \nodata    \\
2M21105881+4657325	&	9.878	&	9.263	&	9.051	&	4	&	24	&	6.87		&	0.39	&	5.40	&	0.80   	&  \nodata    \\
2M21160502+4759308	&	12.483	&	11.812	&	11.418	&	3	&	24	&	$-$12.69	&	0.57	&	21.10	&	1.30   	&  \nodata    \\
2M21202943+4843460	&	11.798	&	11.149	&	10.960	&	4	&	24	&	$-$51.35	&	0.26	&	9.10	&	0.70   	&  \nodata    \\
2M21203355+4554475	&	12.574	&	11.999	&	11.779	&	2	&	12	&	1.70		&  \nodata	&	$<4.00$	& \nodata   &  \nodata    \\
2M21274776+4459041	&	11.585	&	11.011	&	10.780	&	2	&	12	&	$-$18.59	&  \nodata	&	$<4.00$	& \nodata   &  \nodata    \\
2M21311504+1219444	&	10.777	&	10.210	&	9.952	&	10	&	24	&  \nodata	    &  \nodata	&  \nodata &  \nodata  	&  \nodata     \\
2M21323014+4424296	&	12.300	&	11.705	&	11.472	&	2	&	12	&	$-$53.77	&  \nodata	& 	$<4.00$	& \nodata 	&  \nodata      \\
2M21335234+1150104	&	11.935	&	11.351	&	11.083	&	12	&	24	&  \nodata      &  \nodata	&   \nodata &  \nodata 	&  \nodata      \\
2M21351561+4609365	&	12.110	&	11.553	&	11.352	&	2	&	12	&	$-$65.02	&  \nodata	&	5.90	&	0.30   	&  \nodata    \\
2M21353562+0009334	&	12.462	&	11.796	&	11.548	&	3	&	3	&	12.33		&	0.04	&	6.90	&	1.20   	&  \nodata    \\
2M21442066+4211363	&	11.537	&	10.898	&	10.664	&	2	&	12	&	\nodata 	&  \nodata	& \nodata	&  \nodata 	&  SB2  \\
2M21451241+4225454	&	10.936	&	10.328	&	10.085	&	2	&	12	&	$-$59.84	&  \nodata	&	27.30	&  \nodata 	&  \nodata      \\
2M22192066+5818280	&	12.164	&	11.658	&	11.361	&	3	&	3	&	7.31		&	0.22	&	8.90	&	0.60   	&  \nodata    \\
2M22284859+5738281	&	12.336	&	11.725	&	11.449	&	3	&	3	&	13.10		&	0.07	&	8.10	&	2.90   	&  \nodata    \\
2M22392577+5719249	&	8.308	&	7.614	&	7.517	&	3	&	3	&	$-$0.04		&	0.20	&	19.50	&	0.20   	&  \nodata    \\
2M23070962+4647099	&	10.257	&	9.699	&	9.450	&	1	&	3	&	6.04		&  \nodata	&	9.30	&  \nodata 	&  \nodata      \\
2M23081274+4609340	&	11.783	&	11.233	&	10.956	&	1	&	3	&	$-$19.42	&  \nodata	&	12.70	&  \nodata 	&  \nodata      \\
2M23082543+4813393	&	11.098	&	10.425	&	10.163	&	1	&	3	&	$-$33.77	&  \nodata	&	17.20	&  \nodata 	&  \nodata      \\
2M23082777+4700410	&	10.776	&	10.227	&	10.004	&	1	&	3	&	$-$46.80	&  \nodata	& 	$<4.00$	& \nodata 	&  \nodata      \\
2M23110137+4653414	&	11.090	&	10.494	&	10.214	&	1	&	3	&	48.14		&  \nodata	&	6.40	&  \nodata 	&  \nodata      \\
2M23124910+4726557	&	7.498	&	7.192	&	7.127	&	1	&	3	&	3.15		&  \nodata	& 	$<4.00$	& \nodata 	&  \nodata      \\
\end{longtable}
\end{center}

\newpage
\begin{deluxetable}{cccc}
\tablewidth{0pc}
\tabletypesize{\scriptsize}
\tablecaption{Standard Stars Observed During the First Year of the APOGEE M Dwarf Survey.\label{table2}}
\tablecolumns{3}
\tablehead{
\colhead{Star} & \colhead{Alt. Name} & 
\colhead{Standard Type} & \colhead{Reference}
}
\startdata
02085359+4926565	& GJ3136     & vsini     & (a) \\
05030563+2122362	& \nodata    & vsini     & (b) \\
05470907-0512106	& GJ3366     & vsini     & (c) \\
14264827-0510400	& HD126614   & RV        & (d) \\
19051739+4507161	& LHS 3429   & activity  & (e) \\
19121128+4316106	& \nodata    & activity  & (e) \\
19125504+4239370	& \nodata    & activity  & (e) \\
19185898+3812236	& \nodata    & activity  & (f) \\
19235494+3834587	& \nodata    & activity  & (f) \\
19241533+3638089	& \nodata    & activity  & (f) \\
19265550+3844381	& \nodata    & activity  & (f) \\
19271763+3913024	& \nodata    & activity  & (f) \\
19294258+3733223	& LHS6351    & activity  & (e) \\
19302029+3723437	& \nodata    & activity  & (f) \\
19330692+4005066	& \nodata    & activity  & (f) \\
19332454+4515045	& \nodata    & activity  & (e) \\
19333940+3931372	& \nodata    & activity  & (e) \\
19395886+3950530	& \nodata    & activity  & (f) \\
19420033+4038302	& \nodata    & activity  & (f) \\
19445931+4812415	& \nodata    & activity  & (f) \\
19450736+3947341	& \nodata    & activity  & (f) \\
19485718+5015245	& \nodata    & activity  & (e) \\
19510930+4628598	& GJ 1243    & vsini     & (b) \\
22392577+5719249	& HIP 111854 & RV        & (d) \\
23124910+4726557	& LTT 16823  & RV        & (d) \\
\enddata
\tablerefs{ (a) \citep{gizis02}; (b) \citep{nutzman08}; (c) \citep{jenkins09};
(d)\citep{chuback2012}; (e)\cite{ciardi11}; (f)\citep{walkowicz11}
}
\end{deluxetable}

\begin{deluxetable}{ccrcccc}
\tablewidth{0pc}
\tabletypesize{\scriptsize}
\tablecaption{M Dwarf Stellar Properties Derived from Model Fits\label{table3}}
\footnotetext{Emperical low resolution ($\sim$~2000) H-band spectroscopic measurements.}
\footnotetext{Automated APOGEE pipeline values.}
\tablecolumns{7}
\tablehead{
\colhead{Star}    & \colhead{[M/H]}   & \colhead{[M/H]}            & \colhead{T$_{\rm{eff}}$} & \colhead{log g} & \colhead{T$_{\rm{eff}}$} & \colhead{log g}  \\
\colhead{(2MASS)} & \colhead{(Model)} & \colhead{(Observed)$^{a}$} & \colhead{(K)}            & \colhead{(cgs)} & \colhead{(K)$^{b}$}      & \colhead{(cgs)$^{b}$}
}   
\startdata
%  STAR              &  [M/H]            &  Teff           & LOG G         \\
%==================================================================================================================================================== 
00251602+5422547   & $0.0 \pm 0.25$ &  --               & $2700 \pm 100$  & $5.0$  & $2700$ & $4.0$ \\
04125880+5236421   & $0.3 \pm 0.26$ & $-0.04 \pm 0.12$  & $3000 \pm 100$  & $5.0$  & $3300$ & $5.0$ \\
13451104+2852012   & $0.3 \pm 0.20$ & $-0.13 \pm 0.12$  & $3200 \pm 100$  & $5.0$  & $3300$ & $5.0$ \\
19081153+2839105   & $0.0 \pm 0.20$ & $0.22 \pm 0.12$   & $3200 \pm 100$  & $5.0$  & $3300$ & $5.0$\\
\enddata
\end{deluxetable}

\begin{deluxetable}{ccccc}
\tablewidth{0pc}
\tabletypesize{\footnotesize}
\tablecaption{M Dwarf Stellar Properties Derived from Model Fits\label{table4}}
\tablecolumns{5}
\tablehead{
\colhead{Star}    & \colhead{$\rho$}   & \colhead{PA}      & \colhead{$\delta$mag} & \colhead{Instrument}  \\
\colhead{(2MASS)} & \colhead{(mas)}    & \colhead{(degrees)} & \colhead{(Mag)}            & \colhead{} 
}   
\startdata
%===========================================================================================================================

03305473+7041145   & $367.8 \pm 0.2$ & $141.2 \pm 0.2$  & $1.05 \pm 0.05$ & NIRC2  \\
03425325+2326495   & $524.7 \pm 0.2$ & $331.9 \pm 0.2$  & $0.24 \pm 0.03$ & NIRC2  \\
12122940+3940281   & $470.2 \pm 2.0$ & $340.0 \pm 1.0$  & $0.76 \pm 0.06$ & PHARO  \\
\enddata
\end{deluxetable}

\end{document}